\documentclass[prd,aps,reprint,floatfix,superscriptaddress,nofootinbib,preprintnumbers]{revtex4-1}

\usepackage[english]{babel}
\usepackage[utf8x]{inputenc}
\usepackage[T1]{fontenc}

\usepackage{amsmath}
\usepackage{graphicx}
\usepackage{adjustbox}
\usepackage[colorinlistoftodos]{todonotes}
\usepackage[colorlinks=true, allcolors=blue]{hyperref}
\usepackage{subfigure}
\usepackage[toc,page]{appendix}

\newcommand{\apjs}{Astrophys. J. Suppl. Ser.}

\newcommand{\mnras}{Mon. Not. R. Astron. Soc.}

\newcommand{\jcap}{Journal of Cosmology and Astroparticle Physics}
\newcommand{\apjl}{The Astrophysical Journal Letters}

\begin{document}

\title{Gravitational Lensing and the Power Spectrum of Dark Matter Substructure:\\ Insights from the ETHOS $N$-body Simulations}

\author{Ana Díaz Rivero}
\affiliation{Department of Physics, Harvard University, 17 Oxford St., Cambridge, MA 02138}

\author{Cora Dvorkin}
\affiliation{Department of Physics, Harvard University, 17 Oxford St., Cambridge, MA 02138}

\author{Francis-Yan Cyr-Racine}
\affiliation{Department of Physics, Harvard University, 17 Oxford St., Cambridge, MA 02138}
\affiliation{Department of Physics and Astronomy, University of New Mexico, 1919 Lomas Blvd NE, Albuquerque, NM 87131}

\author{Jesús Zavala}
\affiliation{Center for Astrophysics and Cosmology, Science Institute,
University of Iceland, Dunhagi 5, 107 Reykjavik, Iceland}

\author{Mark Vogelsberger}
\affiliation{Department of Physics, Massachusetts Institute of Technology, Cambridge, MA 02139, USA}

%%%%%%%%%%%%%%%%%%%%%%%%%%%%%%%%%%%%%%%%%%%%%%%%%%%%
\begin{abstract}
\noindent Strong gravitational lensing has been identified as a promising astrophysical probe to study the particle nature of dark matter. In this paper we present a detailed study of the power spectrum of the projected mass density (convergence) field of substructure in a Milky Way-sized halo. This power spectrum has been suggested as a key observable that can be extracted from strongly lensed images and yield important clues about the matter distribution within the lens galaxy. We use two different $N$-body simulations from the ETHOS framework: one with cold dark matter and another with self-interacting dark matter and a cutoff in the initial power spectrum. Despite earlier works that identified $ k \gtrsim 100$ kpc$^{-1}$ as the most promising scales to learn about the particle nature of dark matter we find that even at lower wavenumbers - which are actually within reach of observations in the near future -  we can gain important information about dark matter. Comparing the amplitude and slope of the power spectrum on scales $0.1 \lesssim k/$kpc$^{-1} \lesssim 10$ from lenses at different redshifts can help us distinguish between cold dark matter and other exotic dark matter scenarios that alter the abundance and central densities of subhalos. Furthermore, by considering the contribution of different mass bins to the power spectrum we find that subhalos in the mass range $10^7 - 10^8$ M$_{\odot}$ are on average the largest contributors to the power spectrum signal on scales $2 \lesssim k/$kpc$^{-1} \lesssim 15$, despite the numerous subhalos with masses $> 10^8$ M$_{\odot}$ in a typical lens galaxy. Finally, by comparing the power spectra obtained from the subhalo catalogs to those from the particle data in the simulation snapshots we find that the seemingly-too-simple halo model is in fact a fairly good approximation to the much more complex array of substructure in the lens. 
\end{abstract}

\maketitle

\section{Introduction}

Following the many successes of the $\Lambda$-Cold Dark Matter (CDM) standard cosmological model at explaining the universe we observe on large scales \cite{Riess:2006fw,Astier:2005qq,2011ApJS..192...18K,Eisenstein:2005su,Akrami:2018vks,2014MNRAS.444.1518V,2014Natur.509..177V,2018MNRAS.475..676S}, many astrophysicists have turned their sights to subgalactic scales as a way of either reaffirming or falsifying it (see e.g.~Refs.~\cite{Carlberg:2009ae,Carlberg:2011xj,Carlberg:2013eya,2014ApJ...788..181N,2017MNRAS.466..628B,2016MNRAS.463..102E,2016MNRAS.457.3817S,Erkal:2014tda,Erkal:2015kqa,Buschmann:2017ams,Banik:2018pjp,VanTilburg:2018ykj}). Observing these small-scale modes is complicated by the fact that they are deep in the nonlinear regime at low redshifts, and that baryonic effects are generally important for their dynamical evolution. Ironically, star formation becomes increasingly inefficient as halo mass is decreased, meaning that low-mass halos can be difficult - or impossible - to directly observe within the Local Group \cite{Dooley:2016xkj}. Furthermore, different dark matter scenarios that behave like CDM on large cosmological scales can have observable effects on subgalactic scales. Exotic dark matter physics at early times can suppress the formation of low-mass halos (see e.g.~Refs.~\cite{Bond:1983hb,Dalcanton:2000hn,Bode:2000gq,Boyanovsky:2008ab,Boyanovsky:2011aa,1992ApJ...398...43C,Boehm:2001hm,Ackerman:2008gi,Feng:2009mn,Kaplan:2009de,Aarssen:2012fx,Cyr-Racine:2013ab,Chu:2014lja,Buen-Abad:2015ova,Chacko:2016kgg,Ko:2016uft,Krall:2017xcw,Buckley:2014hja,Schewtschenko:2014fca,Cyr-Racine:2015ihg,Agrawal:2017rvu,Sigurdson:2004zp,Dvorkin:2013cea,Xu:2018efh,Gluscevic:2017ywp,Slatyer:2018aqg}) while exotic dark matter physics at late times can change the density profiles of halos \cite{Spergel_1999,Yoshida00,Dave01,Colin02,Vogels_2012,Rocha_2012,Peter:2012jh,Zavala_2013,Kaplinghat:2013xca,Vogelsberger:2015gpr,Kaplinghat:2015aga}. Precisely because of this, probing small-scale structure has become one of the most promising ways of deciphering the particle nature of dark matter. Gaining insight on the distribution and abundance of dark matter substructure in galactic halos can be used to check for consistency with predictions of the CDM paradigm and, if falsified, it can offer clues as to what exotic microphysical properties it might have. 

In recent years, the idea of using strong gravitational lensing to probe substructure has gained traction. Methods such as flux-ratio anomalies in strongly-lensed quasars \cite{Mao:1997ek,Metcalf2001,Chiba:aa,Metcalf:ac,Dalal:aa,Keeton:2003ab,Metcalf:2010aa,Dalal:2002aa,Nierenberg_2014,Nierenberg_2017}, gravitational imaging \cite{Koopmans_2005,Vegetti:2008aa,Vegetti_2010_1,Vegetti_2010_2,Vegetti_2012} and spatially-resolved spectroscopy \cite{Moustakas_Metcalf_2002,Hezaveh_2013,Hezaveh_2016_2,2017MNRAS.472..129A} have been used to either directly detect individual subhalo in lens galaxies, or measure the fraction of mass in substructure within the Einstein radius. These measurements can then be combined to put constraints on the subhalo mass function (see, e.g.~Refs.~\cite{Vegetti:2009aa,Vegetti2014,Li:2015xpc}).

Furthermore, efforts to constrain the statistical properties of subhalo populations by studying collective perturbations of unresolved subhalos on lensed images are well underway \cite{Fadely_2012,Cyr-Racine:2015jwa,Birrer2017,Brewer2015,Daylan:2017kfh}. This is particularly advantageous because of the prediction from CDM that the subhalo mass function rises sharply towards lower masses. A useful statistics to study the collective behavior of the substructure population is the power spectrum of its projected mass density field. This idea was put forth by Ref.~\cite{Hezaveh_2014} (see also Ref.~\cite{2018MNRAS.474.1762C}) and expanded upon in Ref.~\cite{Rivero:2017mao} to show how dark matter microphysics and statistical properties of the subhalo population can get imprinted on this observable. Building on these results, Ref.~\cite{Brennan:2018jhq} used a semi-analytic galaxy formation model to compute this power spectrum and found broad agreement with the predictions of Ref.~\cite{Rivero:2017mao}. Recently, Ref.~\cite{Bayer:2018vhy} analyzed a strongly-lensed image and put an upper bound on the amplitude of the power spectrum, while Ref.~\cite{Cyr-Racine:2018htu} did an in-depth study into how source and lens properties can affect the observability of the substructure power spectrum in strong lenses. 

Fundamentally, the crucial insight to be gained from considering the convergence power spectrum is the ability to describe the effect of substructure in a language that is closer to what strong lensing observations are directly measuring. Indeed, while substructure lensing is often phrased in terms of the subhalo mass function, gravitational lensing observations primarily constrain the length of the deflection vectors at different positions on the lens plane. Since the power spectrum directly describes on which length scales the substructure contributes most to the deflection field, it allows a more direct connection to the actual observations without introducing an intermediate mass function. For the purpose of using lensing observations to extract information about dark matter physics, it is nevertheless important to connect the power spectrum language to the perhaps more familiar halo model of structure formation, for which predictions for different dark matter theories are more readily available. 

In this paper, we present the first in-depth analysis of the dark matter substructure power spectrum in zoom-in $N$-body simulations of galactic halos at redshifts relevant to galaxy-scale strong lensing. We consider two high-resolution simulations of a Milky Way-sized halo, one in which the simulation particles are modeled as being CDM (namely they only interact gravitationally) and another in which they are allowed to self-interact and a cutoff is imposed in the initial cosmological matter power spectrum. We use these simulations to compute the substructure power spectrum and study its behavior as a function of redshift and of dark matter microphysics. We focus here exclusively on the contribution from the subhalos orbiting the main lens galaxy to avoid complications related to multi-plane lensing. Since the line-of-sight contribution \cite{Keeton:2003aa,Li:2016afu,Despali:2017ksx,Birrer:2016xku} to the power spectrum is unlikely to be correlated with the galactic contribution we study here, it can be computed separately. We leave this calculation to future work. 

This paper is organized as follows. In Section \ref{sec:background} we review the expected structure of the substructure power spectrum, as initially derived in Ref.~\cite{Rivero:2017mao}. In Section \ref{sec:simulations} we present the main features of the simulations used in this paper, and in Section \ref{sec:methods} we introduce our methodology. In Section \ref{sec:results} we present our results and we conclude in Section \ref{sec:discussion}.

\section{The Power Spectrum of Dark Matter Substructure within galaxies}\label{sec:background}

In the paradigm of hierarchical structure formation we expect that large, virialized structures in the universe contain an abundance of gravitationally self-bound substructure on a variety of scales. The properties and distribution of this small-scale structure can be studied using strongly lensed images of distant galaxies and quasars. In the limit that most of the lensing is caused by a single massive galaxy (thin-lens approximation), the relevant quantity is the surface mass density $\Sigma$ of the lens galaxy in units of the critical density for lensing $\Sigma_{\rm crit}$, which is usually referred to as the convergence
\begin{equation}
\kappa(\textbf{r}) \equiv \Sigma(\textbf{r})/\Sigma_{\rm crit},
\end{equation}
where
\begin{equation}
\Sigma_{\rm crit} = \frac{c^2 D_{\rm os}}{4 \pi G D_{\rm ol} D_{\rm ls}}.
\end{equation}
Here, $c$ is the speed of light and $G$ the gravitational constant. $D_{xy}$ for $\{x,y\} = \{\text{o,s,l}\}$ represents the angular diameter distance between the observer (o), source (s) and lens (l). The strong lensing regime occurs when $\kappa $ becomes of order unity.  

The total convergence at a point $\textbf{r}$ on the lens plane can be decomposed as
\begin{equation}
\kappa_{\rm tot}(\textbf{r}) = \kappa_0(\textbf{r}) + \kappa_{\rm sub}(\textbf{r}),
\label{eq:kappa_tot}
\end{equation}
where the first term contains the smooth contribution from the main lens galaxy (including both dark matter and baryons) and the second term that of the substructure. Note that in Eq. \eqref{eq:kappa_tot} we include the mean convergence due to substructure $\bar{\kappa}_{\rm sub}$ in the smooth component $\kappa_0$, meaning that $\langle \kappa_{\rm sub} \rangle = 0$. In this work, we assume that the substructure contribution is dominated by self-bound dark matter objects, but note that baryonic objects such as globular clusters and giant molecular clouds could potentially contribute to the galactic substructure. Under the assumptions of the halo model \cite{Cooray:2002dia}, all dark matter is bound in (approximately) spherical halos, meaning that the $\kappa_{\rm sub}$ term above can be decomposed into a sum of individual contributions from each of the $N_{\rm sub}$ subhalos in the lens galaxy
\begin{equation}
\kappa_{\rm sub}(\textbf{r}) = \sum_{i=1}^{N_{\rm sub}} \kappa_{i}(\textbf{r} - \textbf{r}_i),
\label{eq:kappa_sub}
\end{equation}
where $\kappa_i$ and $\textbf{r}_i$ represent the convergence and two-dimensional position of the $i$th subhalo, respectively. In reality, the substructure within a lens is varied, and all contributions to the convergence that cannot be ascribed to the main lens galaxy do not necessarily come from neatly distinguishable subhalos. We will deal with this issue in Section \ref{sec:methods_part} and \ref{sec:results_part}. 

As shown in Ref.~\cite{Rivero:2017mao}, the power spectrum of the $\kappa_{\rm sub}$ convergence field can be written as the sum of a one-subhalo and two-subhalo contributions
\begin{equation}\label{eq:psub_tot}
P_{\rm sub}(\textbf{k}) = P_{\rm 1sh}(\textbf{k}) + P_{\rm 2sh}(\textbf{k}),
\end{equation}
where 
\begin{align}
P_{\rm 1sh}(\textbf{k}) &= \frac{\bar{\kappa}_{\rm sub} \Sigma_{\rm crit}}{\langle m \rangle} \langle\widetilde{\kappa}(\textbf{k})^2\rangle,
\end{align}
and
\begin{align}\label{eq:2sh_term}
P_{\rm 2sh}(\textbf{k}) &= \left( \frac{\bar{\kappa}_{\rm sub} \Sigma_{\rm crit}}{\langle m \rangle}\right)^2 \langle \widetilde{\kappa}(\textbf{k})\rangle^2 P_{\rm ss}(\textbf{k}).
\end{align}
$\widetilde{\kappa}(\textbf{k})$ is the Fourier transform of the subhalo convergence profile, $P_{\rm ss}(\textbf{k})$ is the Fourier transform of the subhalo spatial two-point correlation function (describing subhalo clustering), and the angular brackets represent an ensemble average over subhalo properties such as their mass, truncation radius, and scale radius.

While the substructure power spectrum given in Eq.~\eqref{eq:psub_tot} is in principle anisotropic due to the complex structure of a typical galaxy, we expect the isotropic (monopole) contribution to dominate the signal within the small region probed by strong lensing. This monopole power spectrum is simply given by
\begin{equation}
P_{\rm sub}(k) = \frac{1}{2\pi} \int_0^{2\pi} P_{\rm sub}(\textbf{k}) \; d\phi,
\end{equation}
where
$\phi$ is the polar angle of the $\textbf{k}$ vector. We will focus on this isotropic contribution in the remainder of this paper.

With these general expressions at hand, we showed in Ref.~\cite{Rivero:2017mao} (see also Refs.~\cite{Keeton:2009aa,Cyr-Racine:2015jwa}) that the low-$k$ amplitude of the one-subhalo term is approximately $P_{\rm 1sh}(k) \approx \bar{\kappa}_{\rm sub}m_{\rm eff}/\Sigma_{\rm crit}$, where $m_{\rm eff} \equiv \langle m^2 \rangle / \langle m \rangle$. This means that the one-subhalo term amplitude encodes information on the subhalo mass function and the abundance of subhalos. Deviation from this constant value can inform us about the size of the largest subhalos within the lens galaxy. On the other hand, the amplitude of the two-subhalo term is approximately $P_{\rm 2sh}(k) \propto \bar{\kappa}_{\rm sub}^2 P_{\rm ss}(k)$. Since $\bar{\kappa}_{\rm sub} \ll 1$, the two-subhalo term is generally subdominant as we will see below, but it can dominate the signal on larger scales due to non-negligible subhalo clustering. 

In Ref.~\cite{Rivero:2017mao} we considered two different subhalo populations: one made of truncated Navarro-Frenk-White (NFW) \cite{Navarro:1996gj,Baltz:2007vq} subhalos, and another made of truncated cored subhalos (specifically using a truncated Burkert profile \cite{Burkert:1995yz}). These two models are interesting since they roughly bracket the range of possibilities for the inner density profiles of low-mass subhalos in a broad range of dark matter theories. The slope of this inner density profile could in principle be extracted from the high-$k$ end of the power spectrum since the population of cored subhalos displays a much steeper power spectrum slope there than the one of the cuspy subhalos. However, such a measurement is likely to require high signal-to-noise interferometric data \cite{Cyr-Racine:2018htu}. 

\section{Simulations}\label{sec:simulations}

The $N$-body simulations used in this work are the ETHOS (Effective Theory of Structure Formation) simulations, originally presented in Ref.~\cite{Vogelsberger:2015gpr}. The goal of the ETHOS project \cite{Cyr-Racine:2015ihg,2018MNRAS.477.2886L} is to understand how the fundamental dark matter microphysics affects structure formation on a broad range of scales. To this end, five different dark matter models were investigated: a cold dark matter (CDM) scenario and four scenarios that explore the dark matter parameter space that includes dark matter-dark radiation (DM-DR) interactions, which are responsible for a primordial cutoff in the power spectrum, and self-interacting dark matter (SIDM), labeled ETHOS1-4 depending on the choice of parameter values. In this work we focus on the CDM simulation together with the ETHOS4 model, which has been chosen to reproduce the observed kinematics and properties of Milky Way (MW) dwarf spheroidals. We refer the reader to Refs.~\cite{Vogelsberger:2015gpr,Cyr-Racine:2015ihg} for more details about the ETHOS4 dark matter model, including the values of the particle physics parameters used in the simulations. 

The simulations are initialized at $z=127$  within a $100 h^{-1}$ Mpc periodic box, from which a MW-sized halo ($1.6 \times 10^{12}$ M$_{\odot}$) is chosen to be resimulated. The parent simulation has $1024^3$ particles, a mass resolution of $7.8 \times 10^7 h^{-1}$ M$_{\odot}$, and a spatial resolution (Plummer-equivalent softening length) of $\epsilon = 2 h^{-1}$ kpc. They are thus able to resolve halos down to $\sim 2.5 \times 10^9 h^{-1}$ M$_{\odot}$ with 32 particles. They re-simulate the MW-sized halo to different resolution levels. For this work we use the highest resolution simulation (level 1), where the dark matter particle mass is $m_{\rm DM} = 2.756 \times 10^4$ M$_{\odot}$, $\epsilon = 72.4$ pc and there are approximately $~4.44 \times 10^8$ high-resolution particles in each zoomed-in simulation. 

The cosmological parameters used in the simulations are: $\Omega_{\rm m} = 0.302$, $\Omega_{\Lambda} = 0.698$, $\Omega_{\rm b} = 0.046$, $h=0.69$, $\sigma_8 = 0.839$ and $n_s = 0.967$. The $\Omega_i$ are the density parameters for matter (m), dark energy ($\Lambda$) and baryons (b). $h$ is defined as $h \equiv H_0/100$, where $H_0$ is the Hubble constant. $\sigma_8$ is the amplitude of fluctuations on a scale of 8$h^{-1}$ Mpc, and $n_s$ the primordial index of scalar fluctuations.

\section{Methods}\label{sec:methods}

We extract the substructure power spectrum from the simulations in two ways. We first do so by using the subhalo catalogs obtained using the SUBFIND algorithm \cite{Springel:2000qu} after applying a friends-of-friends (FoF) halo finder with linking length  $b=0.2$. This procedure yields positions for all the detected subhalos together with several subhalo properties, such as the mass, half-light radius, maximum circular velocity, etc. This method is particularly advantageous because it is easier to compare to theoretical predictions, since it closely matches the notion of substructure in the halo model. Furthermore, it allows us to study novel properties of the convergence power spectrum, such as the contribution and detectability of different mass bins. We also extract the power spectrum directly from the particle data in the simulation snapshots (which we shall henceforth refer to as simulation snapshots for brevity) of the zoomed-in MW-like halo. The advantage of this method is that we do not impose any notion of how a subhalo is defined, meaning that all substructure within the simulated galactic halo is captured.

In our fiducial analysis we use the simulation snapshot (and its derived subhalo catalog) at $z=0.5$ (a typical redshift for a lens galaxy), and place the background source at $z=1.5$. With the cosmological parameter values used in the simulations this yields a critical density for lensing $\Sigma_{\rm crit} = 2.35 \times 10^9$ M$_{\odot}$/kpc$^2$.

\begin{figure*}[t!]
\centering
\begin{subfigure}
\centering
\includegraphics[width=0.49\textwidth]{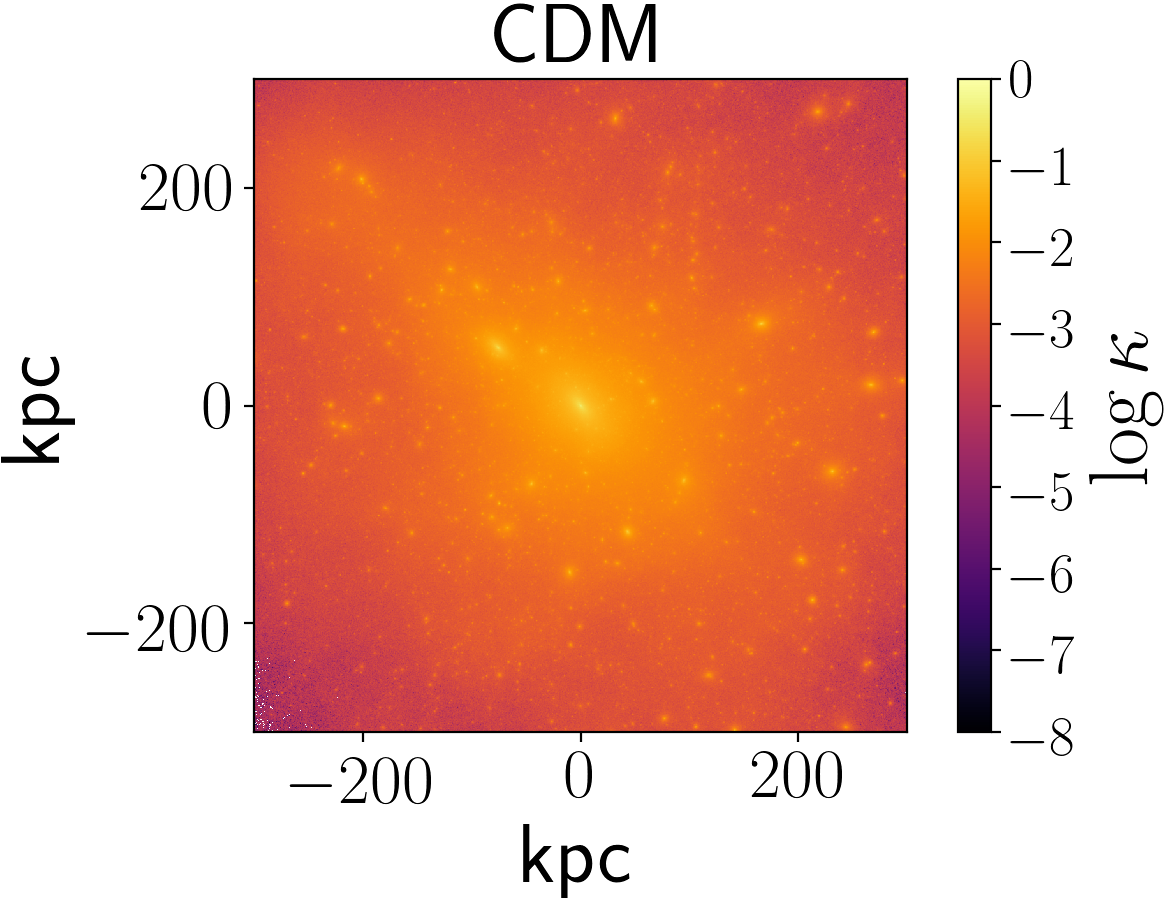}
\end{subfigure}
\begin{subfigure}
\centering
\includegraphics[width=0.49\textwidth]{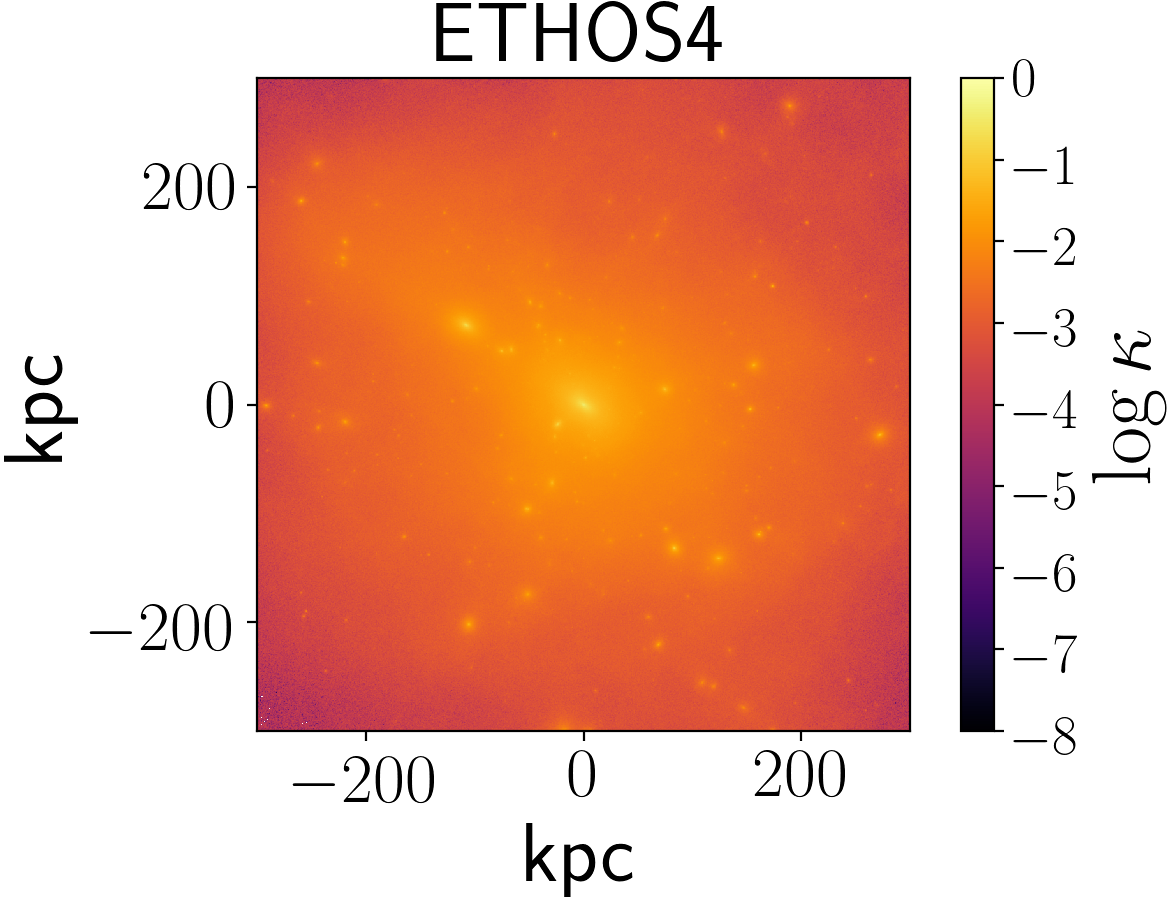}
\end{subfigure}
\begin{subfigure}
\centering
\includegraphics[width=0.49\textwidth]{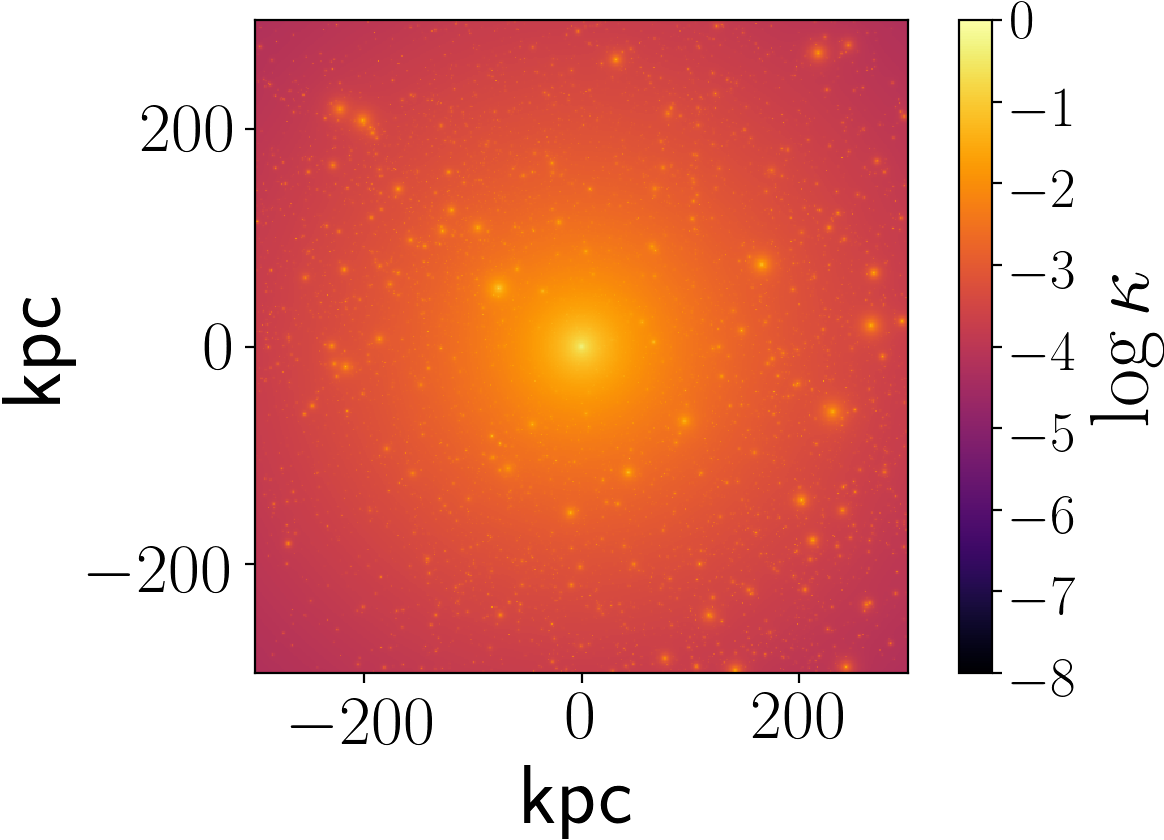}
\end{subfigure}
\begin{subfigure}
\centering
\includegraphics[width=0.49\textwidth]{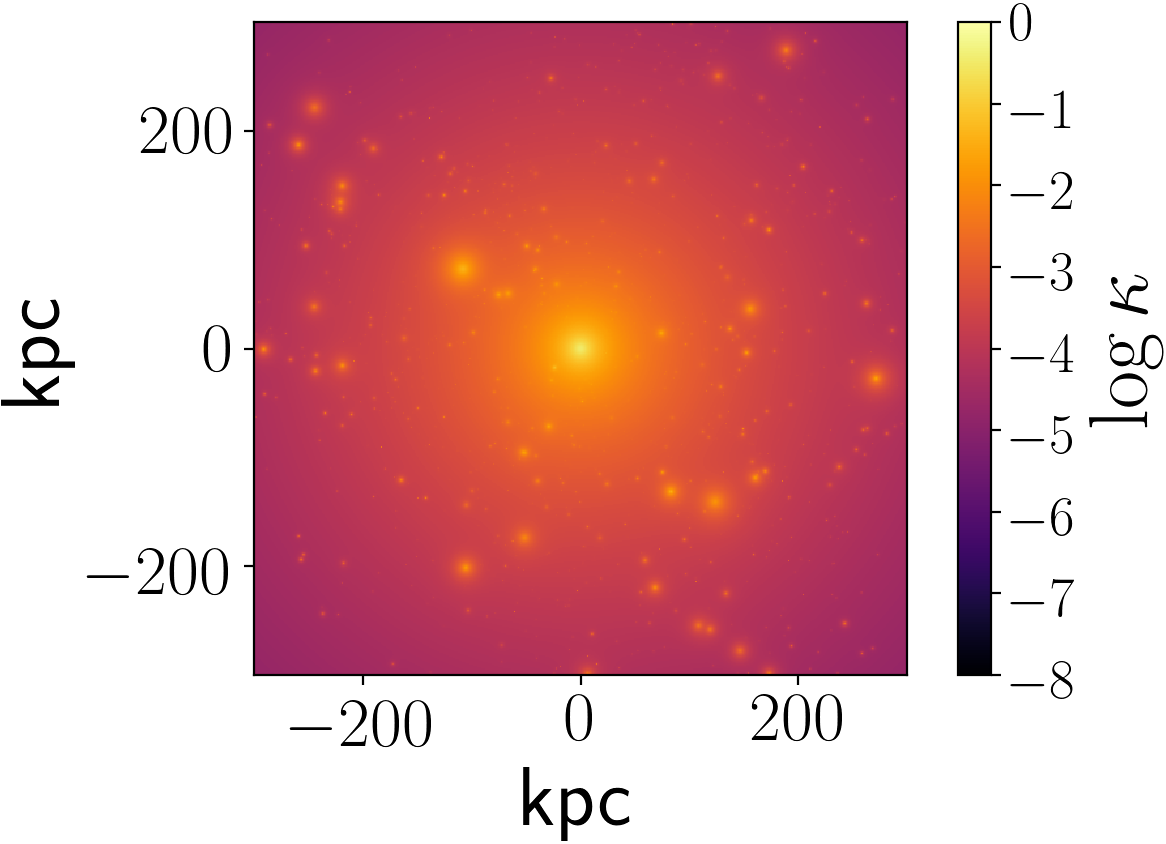}
\end{subfigure}
\begin{subfigure}
\centering
\includegraphics[width=0.49\textwidth]{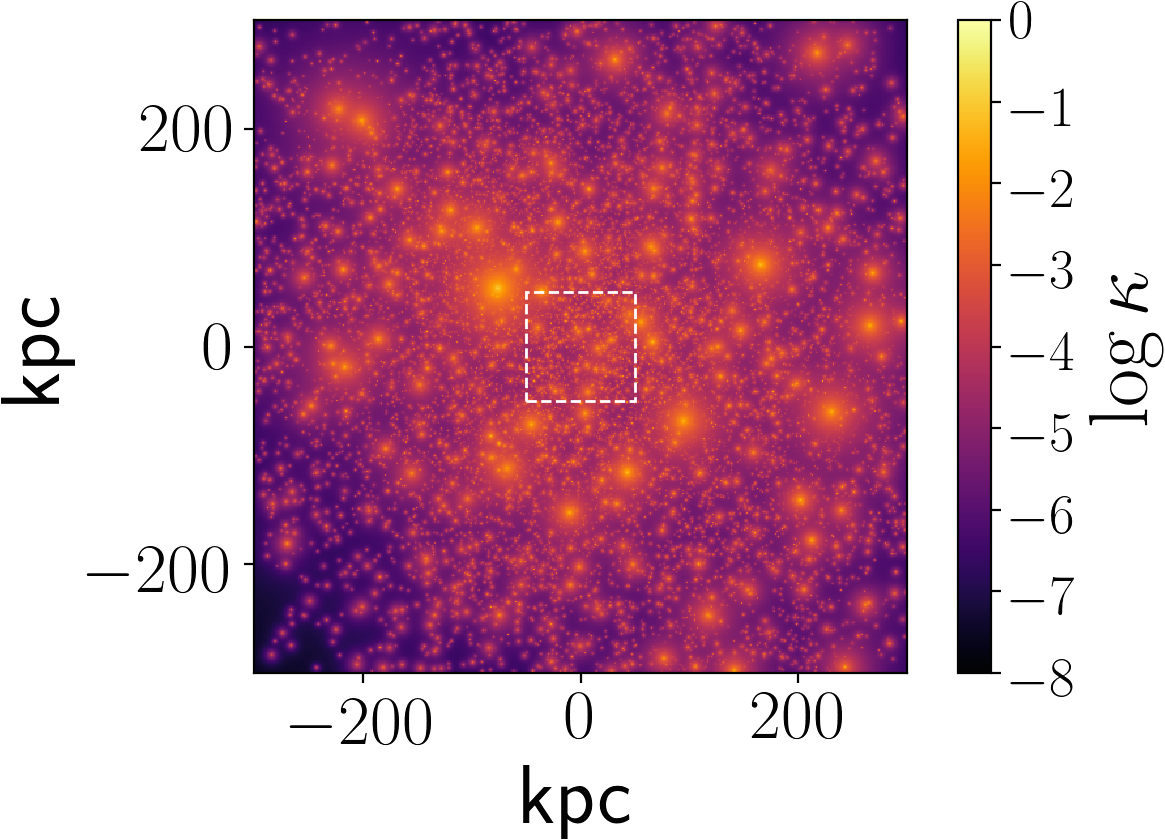}
\end{subfigure}
\begin{subfigure}
\centering
\includegraphics[width=0.49\textwidth]{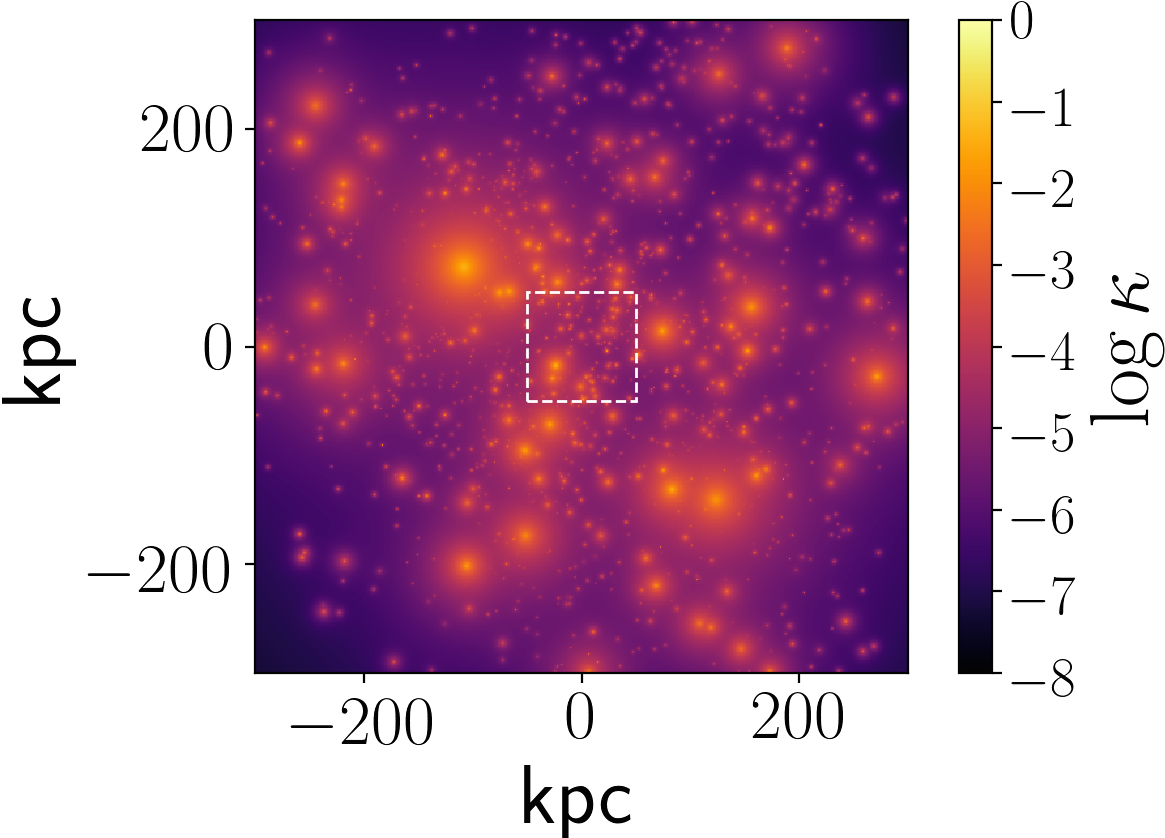}
\end{subfigure}
\caption{\textit{Top left}: convergence field from the particle data for the CDM simulation at $z=0.5$. \textit{Top right}: convergence field from the particle data for the ETHOS4 simulation at $z=0.5$. \textit{Middle left}: convergence field from the subhalo catalog for the CDM simulation at $z=0.5$ with a truncated NFW fit to the host superimposed. \textit{Middle right:} convergence field from the subhalo catalog for the ETHOS4 simulation at $z=0.5$ with a truncated Burkert fit to the host superimposed. \textit{Bottom left}: convergence field from the subhalo catalog for the CDM simulation at $z=0.5$.  \textit{Bottom right}: convergence field from the subhalo catalog for the ETHOS4 simulation at $z=0.5$. The white square in the bottom two panels is centered at the origin and has a size of $L=100$ kpc, therefore it represents the region under consideration in the fiducial case.}\label{fig:projections}
\end{figure*}

Figure \ref{fig:projections} shows the projected density field obtained from the simulation snapshots (top two panels) and built from the subhalo catalogs (bottom two panels) with no mass/resolution threshold imposed. In the middle two panels we have superimposed a host profile on the convergence field obtained from the subhalo catalogs, to serve as a direct comparison to the simulation snapshots. See the section below for details on how this was done. Note that, despite the fact that Milky Way-like halos are generally less massive than typical galaxy-scale gravitational lenses, the two halos we are considering here are not far from being critical, with their convergence fields peaking around 0.3.

\subsection{Power spectra from subhalo catalogs}\label{sec:methods_cat}

We first extract the three-dimensional (3D) subhalo positions from the subhalo catalogs. We only keep subhalos within a comoving cube with side $L=300$ kpc centered on the main lens galaxy, and those that have more than 50 particles, which corresponds to a minimum mass of $1.38 \times 10^6$ M$_{\odot}$. In our fiducial case we limit the highest subhalo mass to $10^8$ M$_{\odot}$, since direct detection methods are expected to be able to detect subhalos above this mass in strong lensing images \cite{2010MNRAS.408.1969V,2012Natur.481..341V,2013ApJ...767....9H}. For completeness, we will also display power spectra that include these more massive subhalos.

To emulate ensemble averaging we project the 3D positions onto $N_{\rm proj}$ different lens planes, which replicates observing different lines-of-sight. We thus end up with $N_{\rm proj}$ two-dimensional (2D) maps of projected positions $\{\textbf{H}_p\}$, where the index $p$ reflects which projection the map corresponds to. We emphasize that considering $N_{\rm proj}$ different projections of the same galaxy can underestimate the variance of the power spectrum: Ref.~\cite{Brennan:2018jhq} compared the variance with 1000 projections of a same subhalo population and that of 1000 independent subhalo populations, and found that the latter was significantly larger. They did however find that the difference between both scenarios was much smaller when the largest subhalos are removed (they imposed $m_{\rm high} = 10^9$ M$_{\odot}$), meaning that for our fiducial case we don't expect to be underestimating the variance so drastically.

Subhalos in the CDM simulations are shown to be well fit by NFW profiles, so we fit a (truncated) NFW convergence profile to each subhalo in the projected map (see Appendix \ref{sect:conv_profiles}). This profile is determined by three subhalo parameters: the total mass $m$, the scale radius $r_{\rm s}$ and the tidal truncation radius $r_{\rm t}$. Note that truncating the NFW profile ensures that each subhalo has a total finite mass. 

The subhalo finder assigns a gravitationally-bound mass to each subhalo, which we identify with the total mass parameter $m$ of a truncated NFW subhalo. We obtain the scale radius of our subhalos using the well-known relation $r_{\rm max}/r_{\rm s} = 2.1626$ \cite{2014MNRAS.444..222G} for the NFW profile, where $r_{\rm max}$ is the radius at which the maximum circular velocity of the subhalo is attained, which the subhalo finder computes. We finally set the value of the tidal radius $r_{\rm t}$ by numerically solving the nonlinear relation $2\,m(< r_{\rm half}) = m$, where $r_{\rm half}$ is the radius containing half the subhalo mass, which is also reported by the subhalo finder.\footnote{We note that we could have simultaneously solved for both $r_{\rm s}$ and $r_{\rm t}$ using the nonlinear relations $2\,m(< r_{\rm half}) = m$ and $(dv^2/dr)|_{r_{\rm max}}=0$, where $v$ is the circular velocity profile of the subhalo. Our tests show that doing so leads to differences in the substructure power spectrum that are smaller than the scatter between different projections.}

Due to the presence of dark matter self-interaction, the subhalos in the ETHOS4 simulation are instead fit with truncated Burkert profiles (see Appendix \ref{sect:conv_profiles}), which can be fully specified by three parameters, namely the total mass $m$, the Burkert radius $r_{\rm b}$, and the tidal truncation radius $r_{\rm t}$. It is useful to write the Burkert radius as $r_{\rm b} = p\,r_{\rm s}$, where $p$ defines the core size. We use a similar procedure as above to obtain the values of $r_{\rm s}$ and $r_{\rm t}$ from the subhalo catalog, fixing $p=0.666$ to ensure that the standard kinematic relation $r_{\rm max}/r_{\rm s} = 2.1626$ is preserved. As a check of our calibration procedure, we compute the predicted values of $v_{\rm max}$ from our Burkert fits and compare those to the corresponding catalog entries, finding at most a $20\%$ scatter between these values.

Notice that, although we have included all the subhalos within a cube with side $L=300$ kpc, strong lensing cannot probe such a large area transverse to the line of sight (LOS). Therefore, after projecting we limit the box size to either $L = 100$ kpc, i.e. $\pm 50$ kpc from the host center, or $L = 200$ kpc, depending on the scales we want to probe. Conversely, strong lens images do give us access to the entire LOS volume of the main halo, which is why it is important to first allow all the subhalos within the host to be projected before limiting the box size transverse to the LOS to a more realistic\footnote{We note that our projected area with sides of comoving length $L=100$ kpc is still larger than a typical galaxy-scale strong lensing region. This allows us to capture the impact of subhalos that are on the outskirts of the strong lensing region but can still influence the lensed images.} observable size. 

Applying this procedure to the 2D position maps $\{\textbf{H}_p\}$, we obtain $N_{\rm proj}$ 2D convergence maps $\{\boldsymbol{\kappa}_p\}$, which we Fourier transform and square to obtain an estimate of the 2D power spectrum for each individual map $|\widetilde{\boldsymbol{\kappa}}_p({\bf k})|^2$. A factor of $A_{\rm pix}^2/A_{\rm box}$ is necessary to normalize each power spectrum, where $A_{\rm pix}$ is the pixel area and $A_{\rm box}$ is the box area. An estimate of the monopole substructure power spectrum $P_{{\rm sub},p}(k)$ from the $p$th convergence map is finally computed by azimuthally averaging $|\widetilde{\boldsymbol{\kappa}}_p({\bf k})|^2$,
\begin{equation}\label{eq:ang_ave_est}
P_{{\rm sub},p}(k) = \frac{1}{2\pi} \int_0^{2\pi} |\widetilde{\boldsymbol{\kappa}}_p({\bf k})|^2 \; d\phi.
\end{equation}
Repeating this procedure for our $N_{\rm proj}$ maps, we can compute the average substructure power spectrum $\bar{P}_{\rm sub}(k)$ as well as the  68th and 90th percentiles characterizing the distribution of power spectrum values at each wavenumber. We generally find that for a given $k$ bin the power spectrum values are not Gaussian-distributed.

Carrying out this procedure we obtain the \textit{total} subhalo power spectrum, including both the one- and two-subhalo contributions (see Eq.~\eqref{eq:psub_tot}). It is however possible to isolate the two-subhalo term by simply carrying out the procedure above directly from the position projections. For each map $\textbf{H}_p$ we create a 2D map $\textbf{S}$ of 
\begin{equation}
\mathbf{S}_{p,j} = \frac{N_{p,j} - \bar{N}_{p}}{\bar{N}_{p}},
\end{equation}
where for the $p$th projection $N_{p,j}$ is the number of subhalos in the $j$th spatial pixel and $\bar{N}_p$ is the average number of subhalos per pixel. We can then follow the same procedure to Fourier transform and azimuthally average to obtain $P_{\rm ss}(k)$. The two-subhalo contribution can then be computed according to Eq.~\eqref{eq:2sh_term} given a choice of the subhalo convergence profile. It is important to avoid over-counting the subhalo clustering, since it contributes both in the $P_{\rm ss}(\textbf{k})$ term and the $\langle \widetilde{\kappa}(\textbf{k})\rangle$ term. To avoid this issue when isolating the two-subhalo term we randomize all the subhalo positions within a given projection before making the convergence maps.

Finally, we want to point out that the smallest $k$ mode accessible is determined by the box size as $k_{\rm min} = 2 \pi / L_{\rm box}$, while the largest $k$ mode accessible is determined by the pixel size, $k_{\rm max} = 2 \pi / L_{\rm pix} = 2 \pi N_{\rm pix}/ L_{\rm box}$, where $N_{\rm pix}$ is the number of pixels on a given size of the box. Unless otherwise mentioned, we limit the box size to $L_{\rm box} = 100$ kpc (symmetrically centered about the host center). For computational efficiency we limit the image resolution to be $501 \times 501$ pixels. Thus, $k_{\rm min} \approx 0.06$ kpc$^{-1}$ and $k_{\rm max} \approx 30$ kpc$^{-1}$.

\subsection{Power spectra from simulation snapshots}\label{sec:methods_part}

The level-1 ETHOS simulations we use in this work contain almost half a billion particles, meaning that it can be quite costly to carry out this analysis at the $N$-body particle level. We use the publicly available code \texttt{nbodykit} \cite{Hand:2017pqn} to perform parts of our analysis. \footnote{\texttt{nbodykit} is an open source large-scale structure toolkit written in Python.}. All its algorithms are parallel, which greatly expedites the analysis procedure. 

Starting from particle catalogs, \texttt{nbodykit} can build a density mesh equal to $1+\delta(\textbf{x})$, meaning that to obtain the convergence field we have to rescale the mesh with factors of the average number density of particles $\bar{n}$, $\Sigma_{\rm crit}$, and the $N$-body particle mass $m_{\rm part}$: 
\begin{equation}
\kappa(\textbf{x}) = \frac{\bar{n} \; m_{\rm part}}{\Sigma_{\rm crit}}(1+\delta(\textbf{x})).
\end{equation}
Much like our analysis based on the subhalo catalogs, we limit the particles out to 300 kpc from the host center, but we do not impose any resolution/mass thresholds for inclusion. 

\begin{figure*}[t!]
\centering
\includegraphics[width=0.85\textwidth]{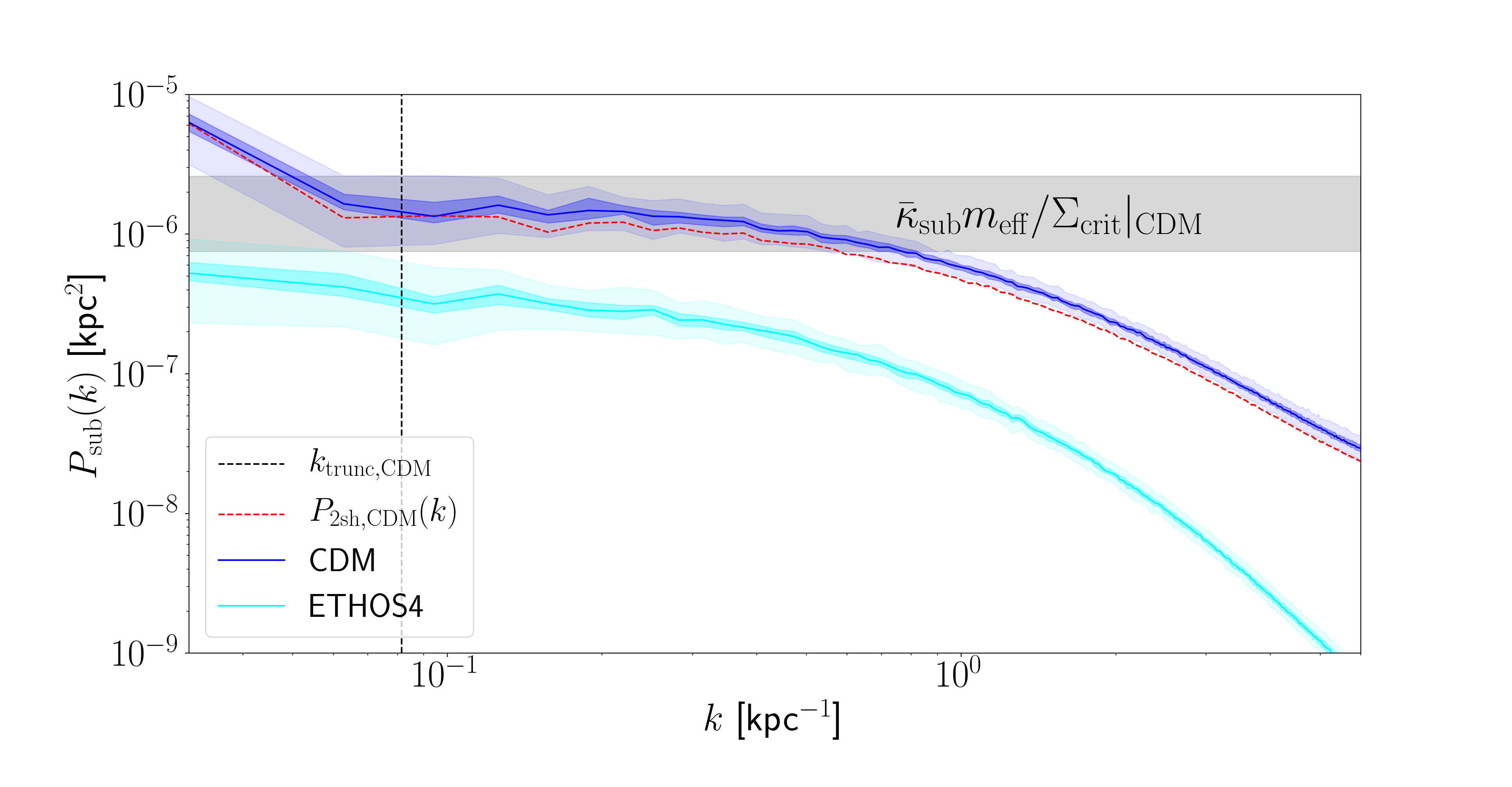}
\caption{Substructure convergence power spectrum from the subhalo catalog at $z=0.5$ and $m_{\rm high} = 10^8$ M$_{\odot}$ for both the CDM (blue) simulation and the ETHOS4 (cyan) simulation for a box with side $L=200$ kpc. The shaded gray horizontal region shows the predicted amplitude from Ref. \cite{Rivero:2017mao} with $\bar{\kappa}_{\rm sub}$ and $m_{\rm eff}$ (and their associated errors) obtained from the CDM simulations, and the vertical dashed line is the median $k_{\rm trunc} \equiv 1/r_{\rm trunc,max}$. The red dashed line is the isolated two-subhalo contribution for the CDM simulation. The wavenumbers $k$ are in comoving coordinates.}\label{fig:cdm_ethos4}
\end{figure*}

Unlike our catalog-based analysis where we were able to isolate the substructure contribution $\kappa_{\rm sub}$ in Eq.~\eqref{eq:kappa_sub}, we instead directly obtain the total convergence $\kappa$ from the simulation snapshot. To isolate the substructure signal we are interested in, we therefore have to subtract the main host halo contribution $\kappa_0$. Our approach to remove this contribution consists of averaging many different projections to approximate the host profile, 
\begin{equation} \label{eq:kappa_host}
\kappa_{\rm host}(\textbf{r}) \approx \langle \kappa_{\rm box}(\textbf{r}) \rangle,
\end{equation}
and then subtracting this average map from a given projection $p$ to obtain our estimate of the 2D substructure power spectrum
\begin{align} \label{eq:pk_box-host}
|\widetilde{\boldsymbol{\kappa}}_p({\bf k})|^2 &= \left[\int d^2\textbf{r} \; e^{-i \textbf{k} \cdot \textbf{r}} \left(\kappa_{{\rm box},p}(\textbf{r}) - \kappa_{\rm host}(\textbf{r})\right)\right]^2,
\end{align}
before performing the angular averaging as in Eq.~\eqref{eq:ang_ave_est}.

Due to the discrete nature of the simulation particles, we impose a conservative $k_{\rm max}$ cut beyond which we do not trust the results. For our choice of box size and $N_{\rm mesh} = 1024$, we impose $k_{\rm max} = 3$ kpc$^{-1}$.

\section{Results}\label{sec:results}

\subsection{Subhalo catalog}\label{sec:results_cat}

\begin{figure*}[t!]
\centering
\begin{subfigure}
\centering
\includegraphics[width=0.49\textwidth]{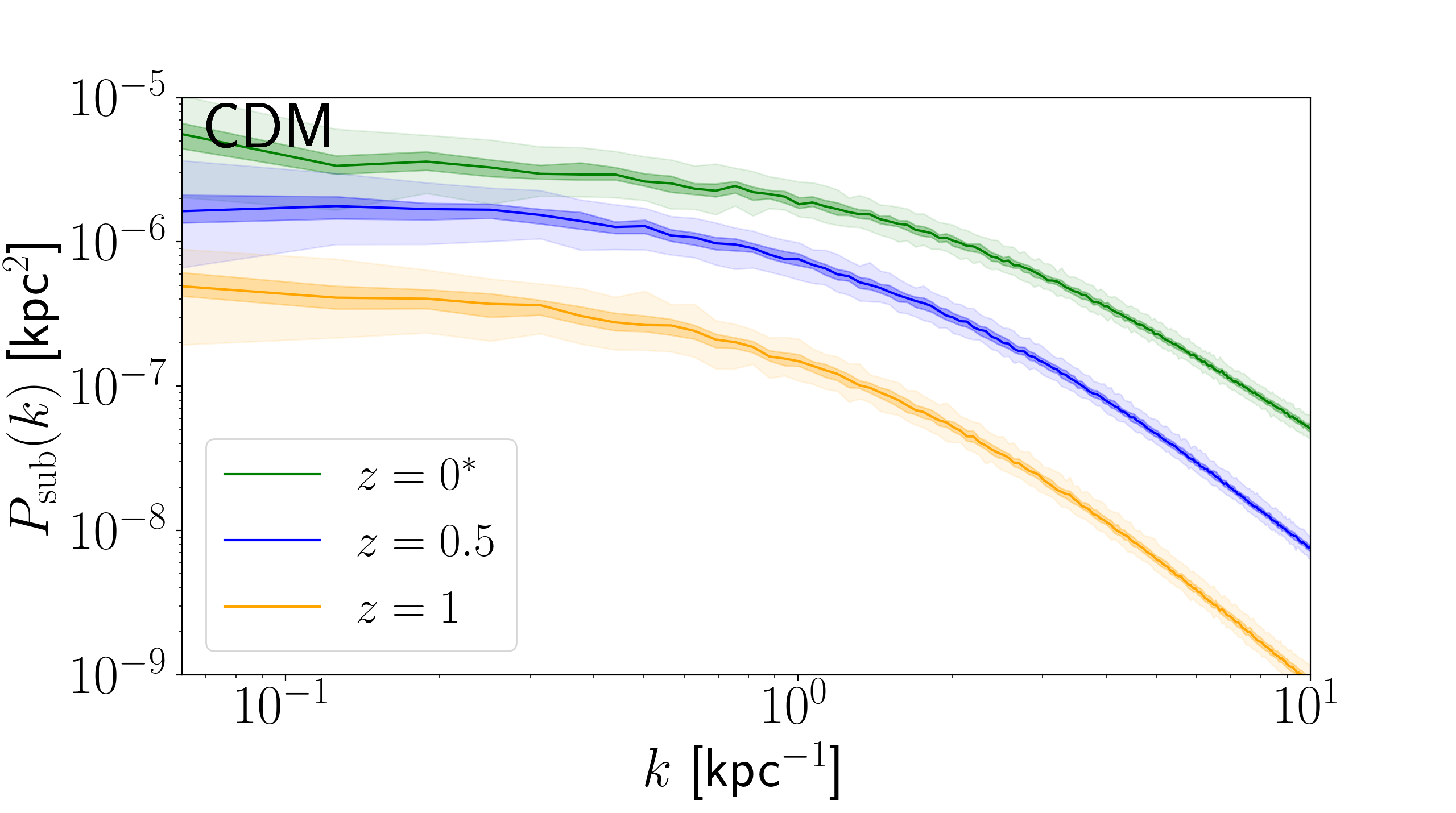}
\end{subfigure}
\begin{subfigure}
\centering
\includegraphics[width=0.49\textwidth]{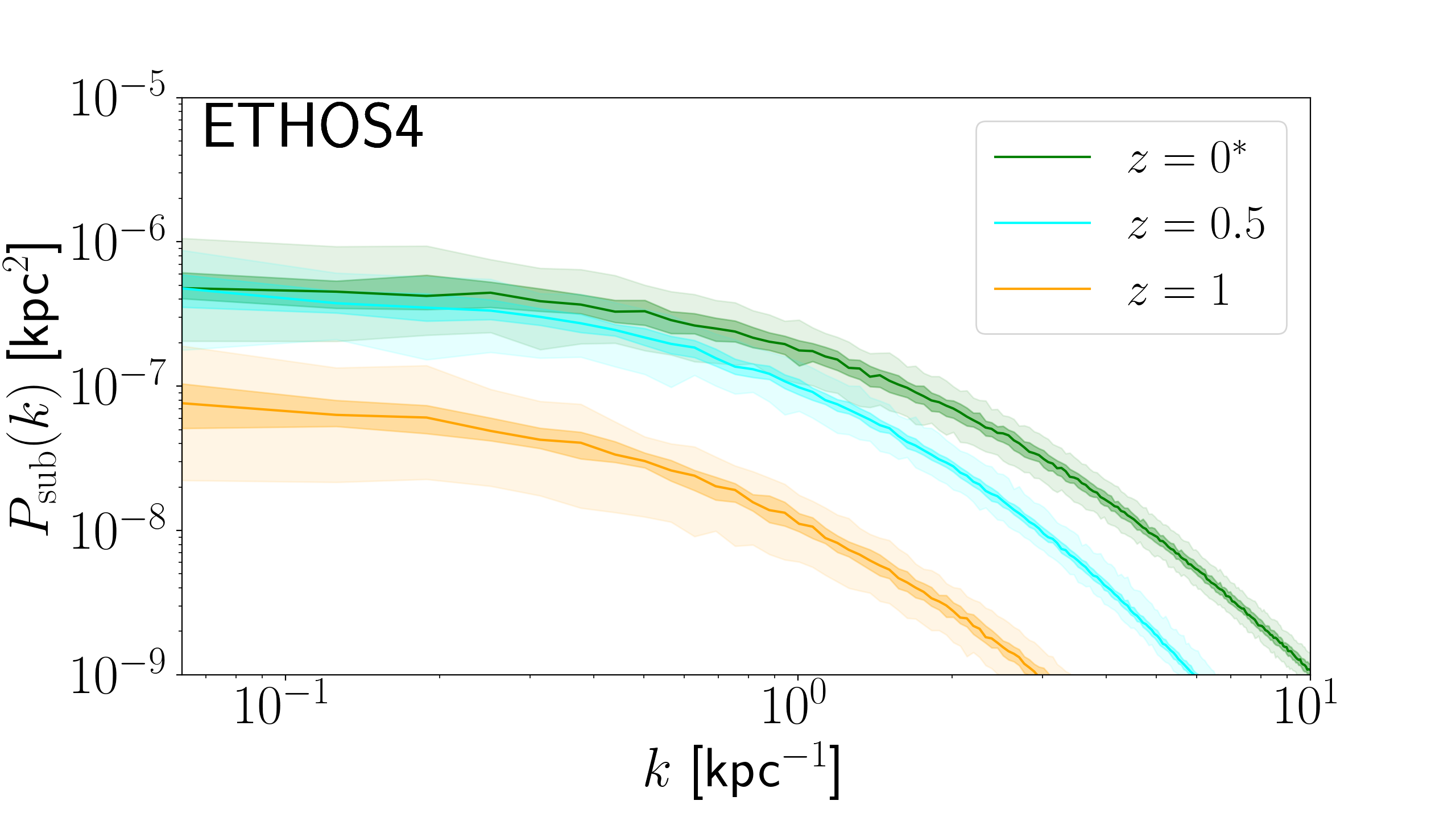}
\end{subfigure}
\begin{subfigure}
\centering
\includegraphics[width=0.49\textwidth]{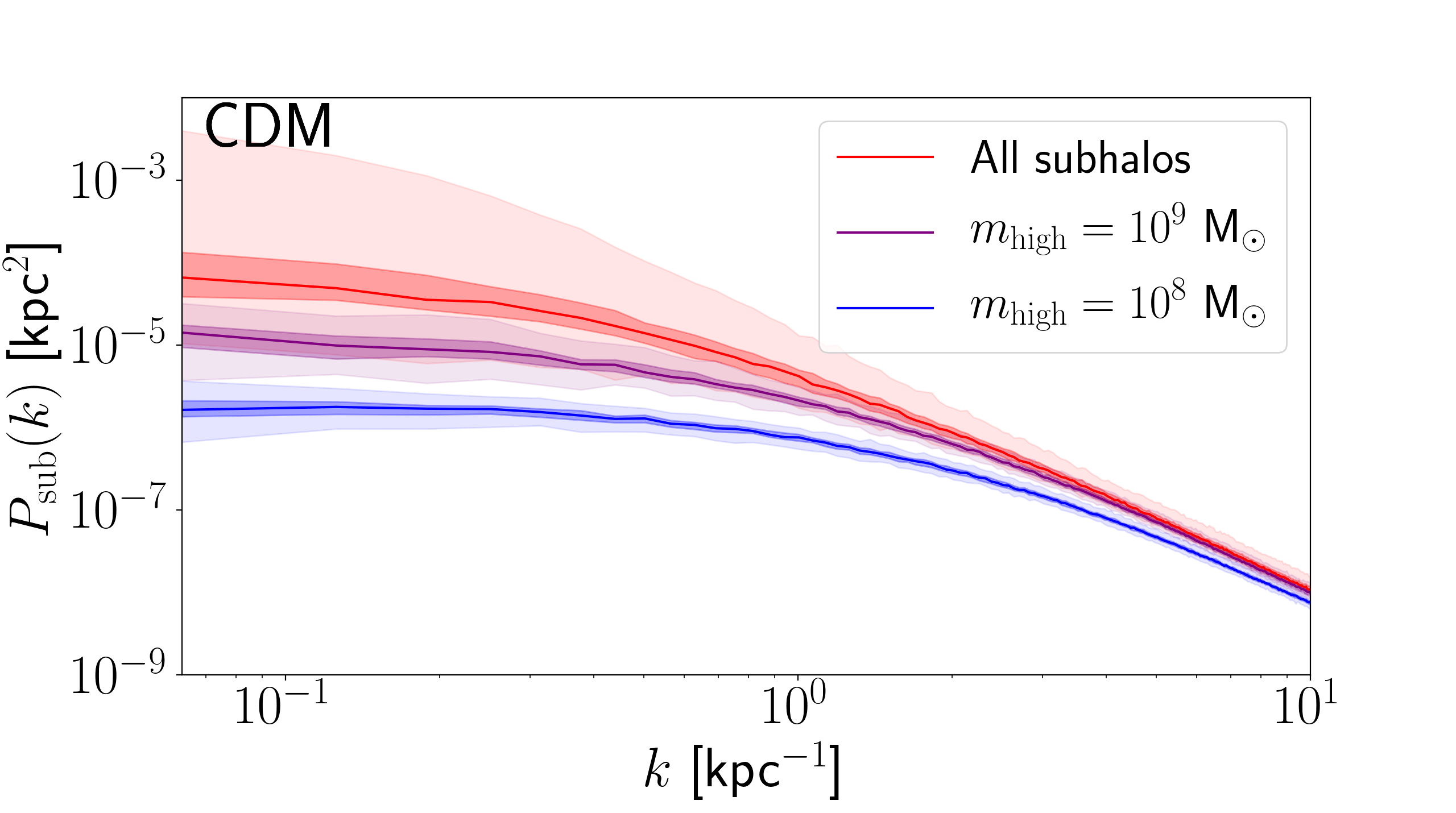}
\end{subfigure}
\begin{subfigure}
\centering
\includegraphics[width=0.49\textwidth]{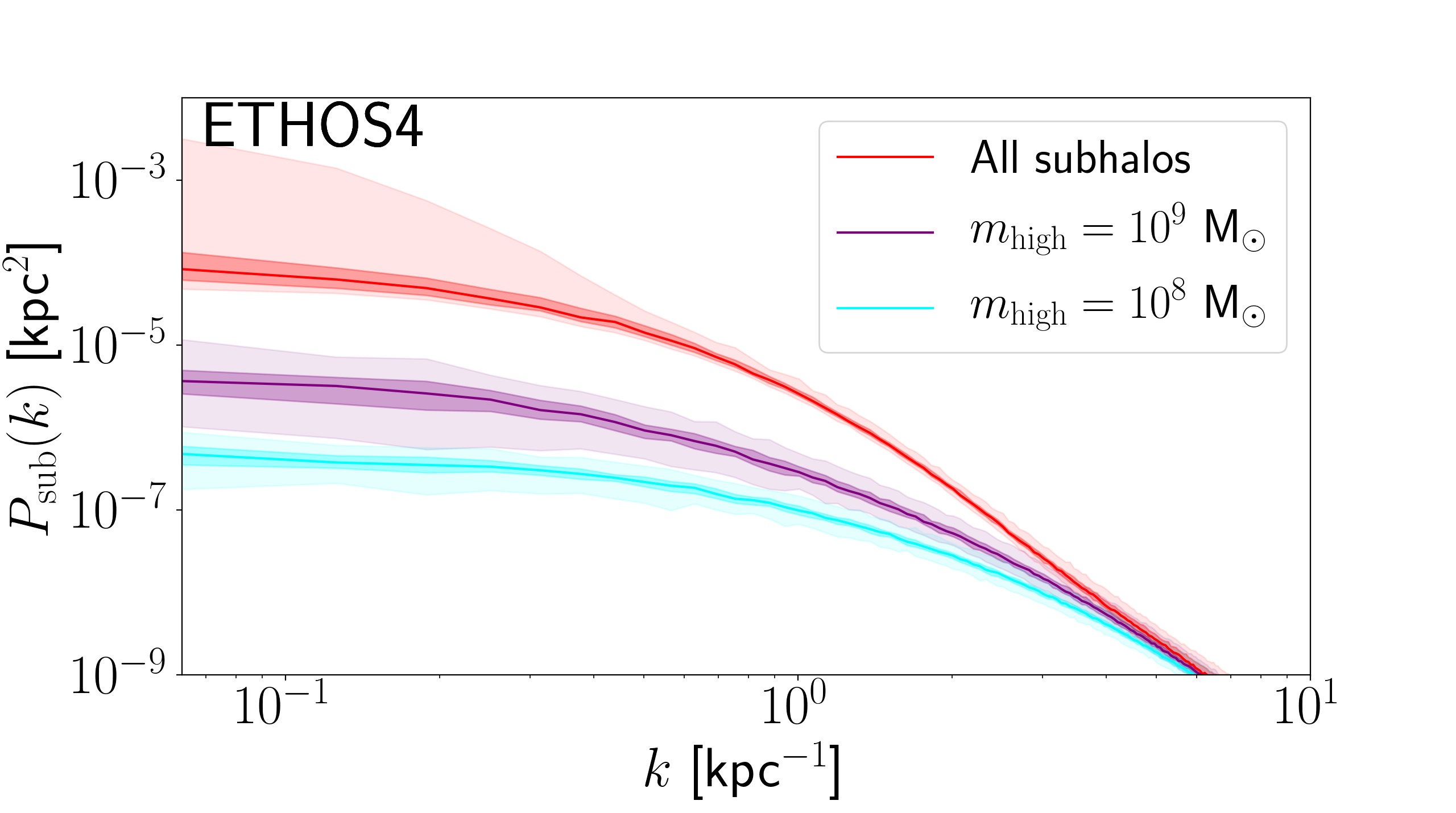}
\end{subfigure}
\caption{\textit{Top Left}: redshift dependence 
of the convergence power spectrum for the CDM simulation. \textit{Top Right}: redshift dependence 
of the convergence power spectrum for the ETHOS4 simulation. \textit{Bottom Left}: mass dependence 
of the convergence power spectrum for the CDM simulation. \textit{Bottom Right}: mass dependence 
of the convergence power spectrum for the ETHOS4 simulation. Note that the $y$-axis is the same for a given row but differs between rows. The wavenumbers $k$ are in comoving coordinates. 
*As discussed in the text, the $z=0$ power spectra are computed using the subhalo catalog at $z=0$ but the distance between the observer and the lens $D_{\rm ol}$ is fixed to be the same as for a lens at $z=0.5$ because $\Sigma_{\rm crit}$ diverges as $z \rightarrow 0$. }\label{fig:cat}
\end{figure*}

Figure \ref{fig:cdm_ethos4} shows the convergence power spectrum for the fiducial CDM (blue) and ETHOS4 (cyan) simulations for the larger box size, $L = 200$ kpc. This larger projected area allows us to be sensitive to the two-subhalo term on sufficiently large scales. For the CDM case, the two-subhalo term appears as an upturn in the power spectrum for $k\lesssim 0.06$ kpc$^{-1}$. To show that this upturn is indeed due to subhalo clustering we have overlaid the isolated two-subhalo contribution in dashed red, obtained with the method outlined in \ref{sec:methods_cat}. As explained in Refs.~\cite{Chamberlain:2014fja,Rivero:2017mao}, this two-subhalo term corresponds to the so-called ``host'' contribution arising because all subhalos are gravitationally bound to their host galaxy. Of course, such small values of $k$ are unobservable with the small field of view of typical strongly lensed images.

This figure has been made with 90 different projections. The solid lines correspond to the median and the shaded regions to the 68\% and 90\% confidence level areas. The vertical dashed line corresponds to the truncation wavenumber for the CDM simulation, defined as $k_{\rm trunc} \equiv 1/r_{\rm t,max}$. As predicted in Ref. \cite{Rivero:2017mao}, the break in the power spectrum is related to the size of the largest subhalos and the two-subhalo term becomes dominant below $k_{\rm trunc}$. We can see that the amplitude of the one-subhalo term is well approximated by $\bar{\kappa}_{\rm sub} m_{\rm eff} / \Sigma_{\rm crit}$ (just as for $k_{\rm trunc}$ we only show this for the CDM simulation for clarity, but the same applies to the power spectrum obtained from the ETHOS4 simulation). The amplitude of the power spectrum is noticeably lower in ETHOS4, since there are many fewer subhalos. This dearth of small-mass subhalos is also responsible for the steeper slope at $k\gtrsim1$ kpc$^{-1}$ in ETHOS4. The power spectrum slope on these scales appears as a key observable that can probe the abundance of small-mass subhalos in lens galaxies. Finally, we can see that the two-subhalo term does not appear to contribute significantly to the ETHOS4 power spectrum on large scales. Indeed, the small overall number of subhalos in this case makes it difficult to probe the subhalo clustering signal.

\begin{figure*}[t!]
\centering
\includegraphics[width=0.47\textwidth]{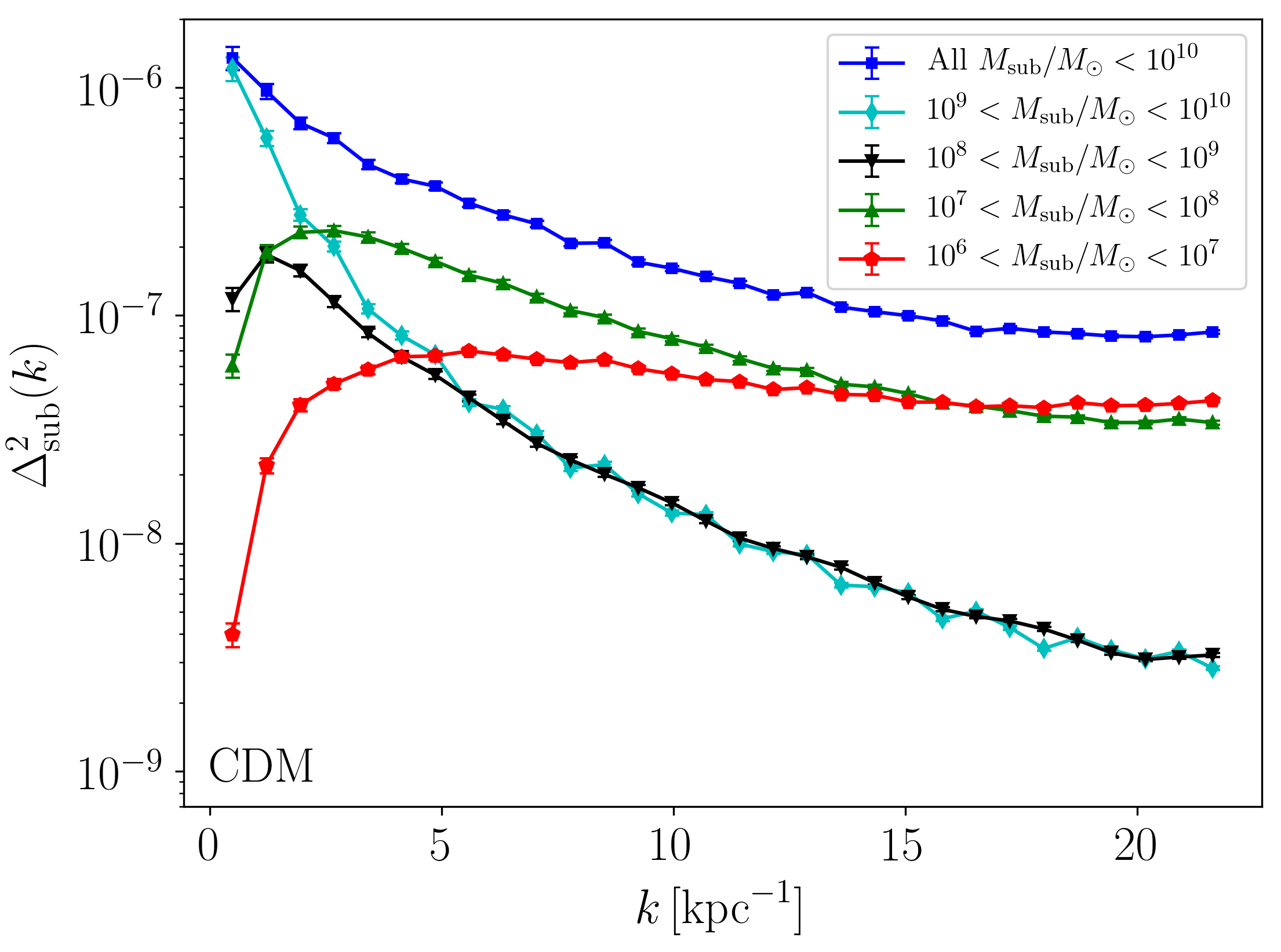}
\includegraphics[width=0.47\textwidth]{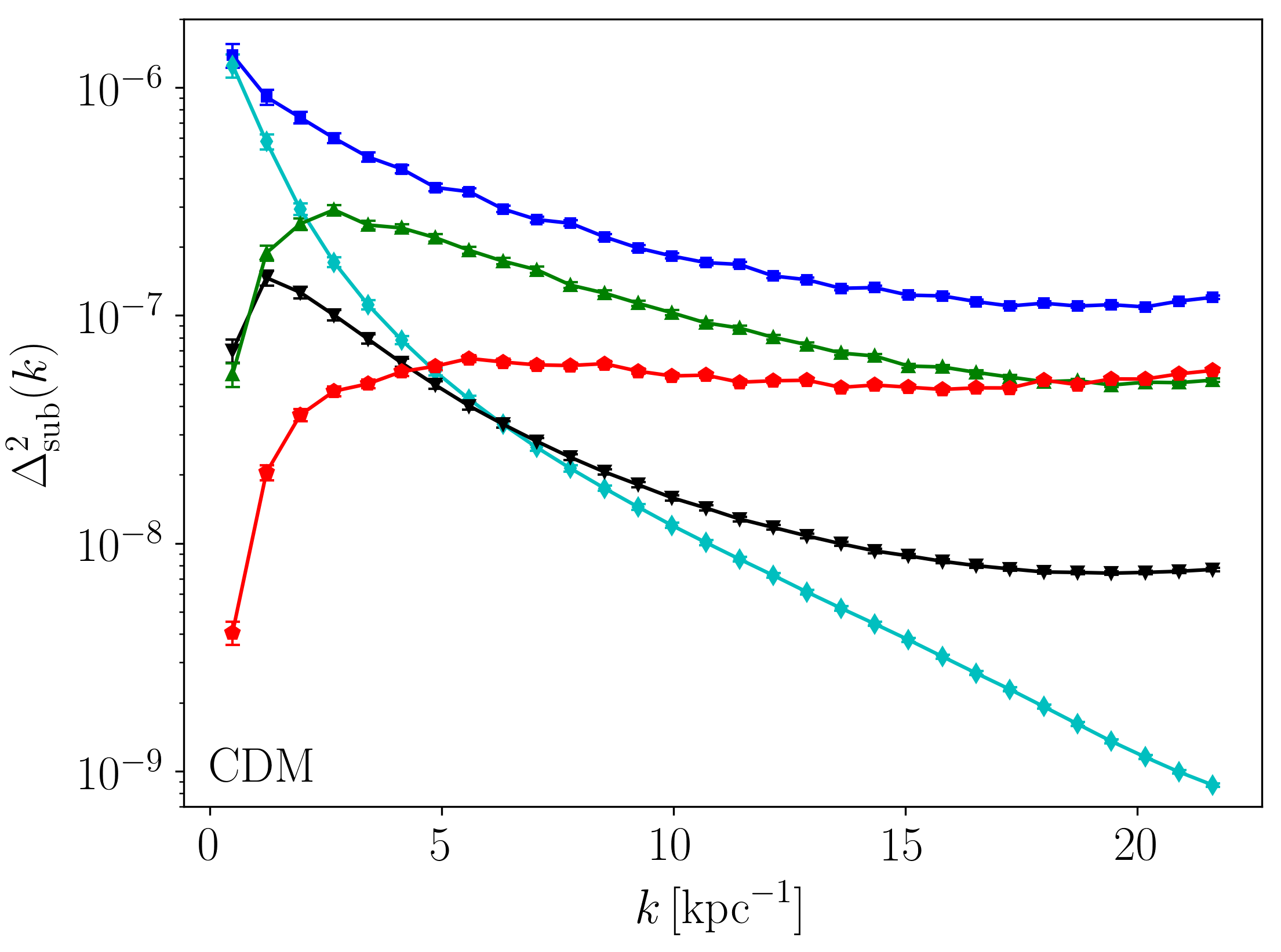}\\
\includegraphics[width=0.47\textwidth]{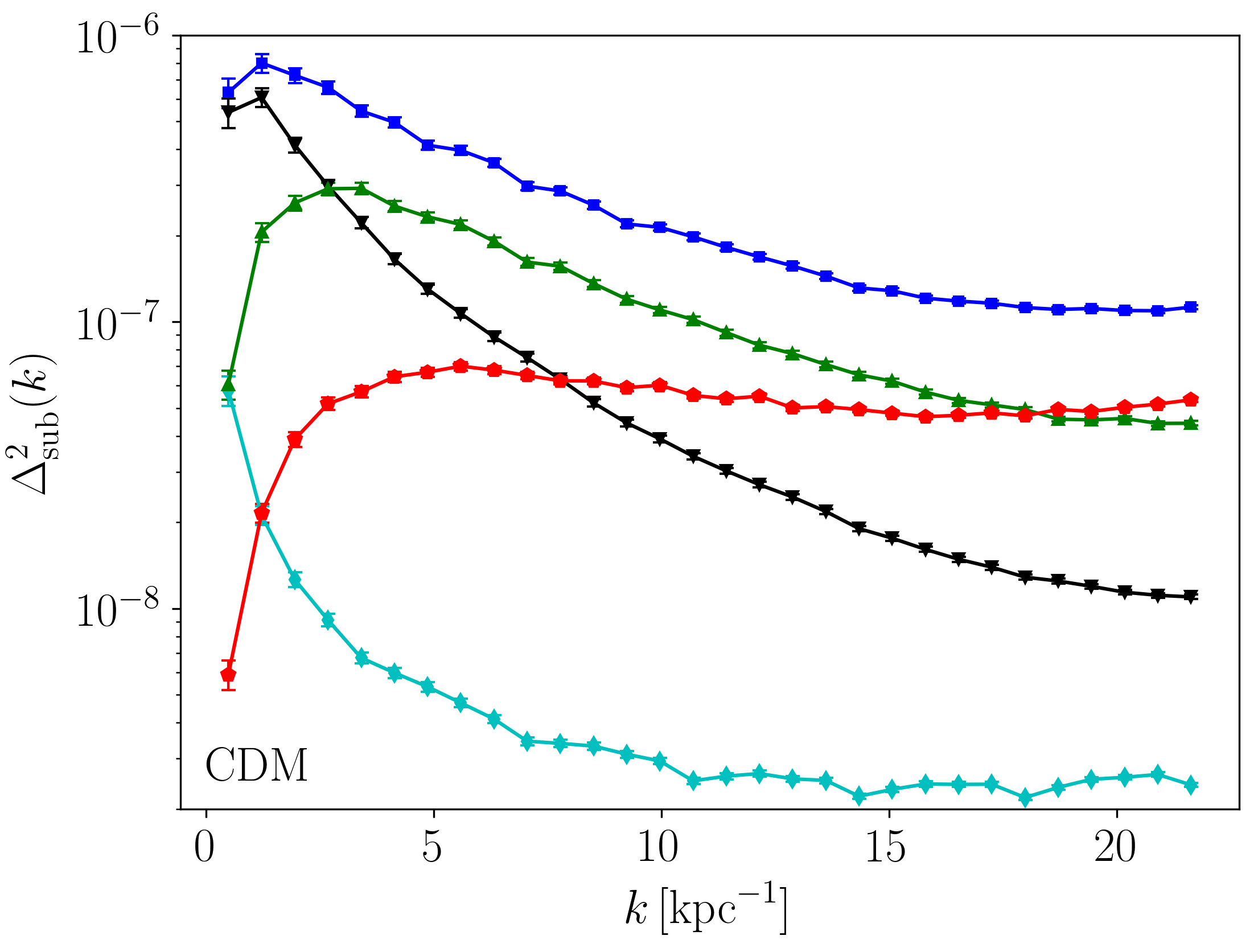}
\includegraphics[width=0.47\textwidth]{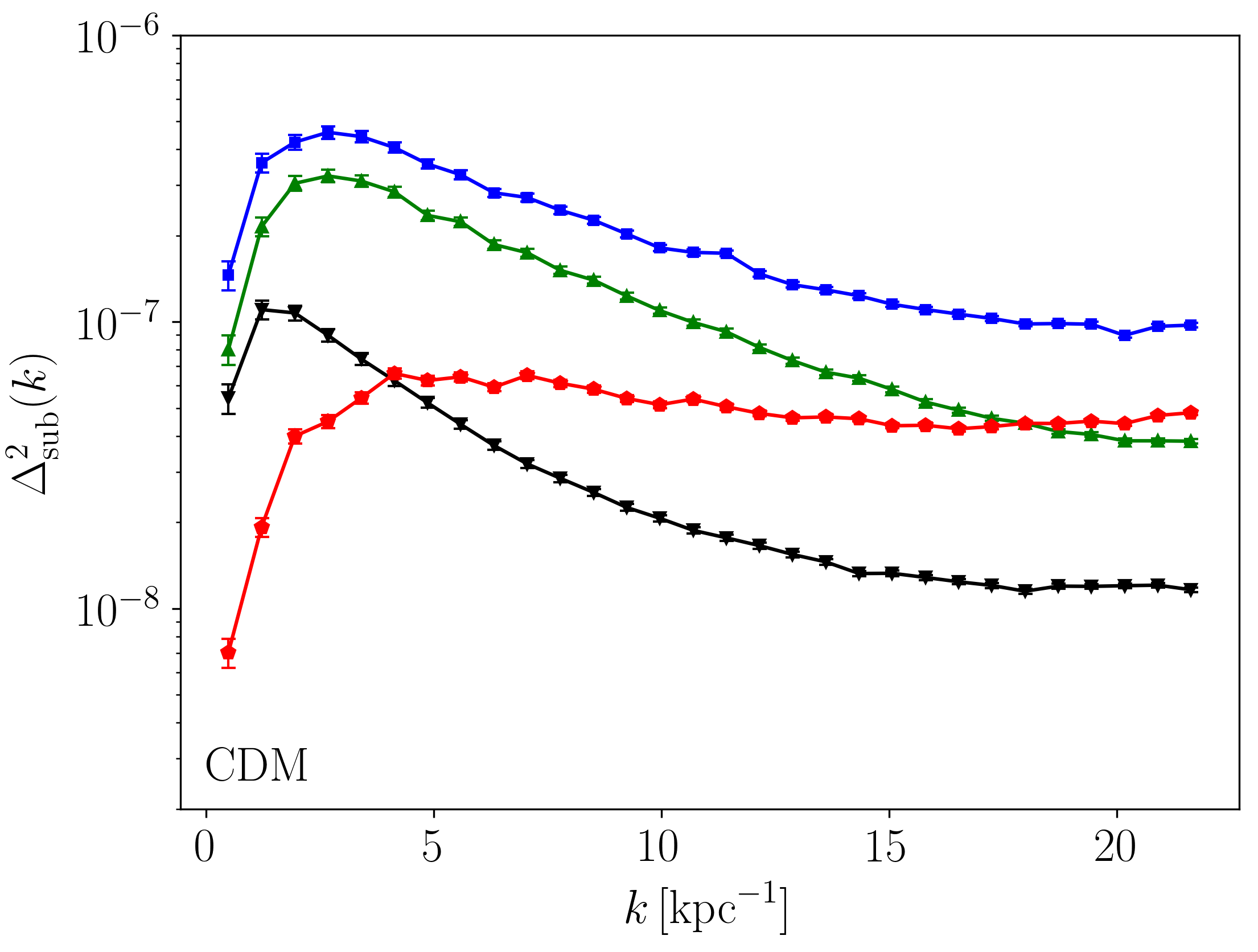}
\caption{Decomposition of the CDM substructure power spectrum into its contributions from subhalos in different mass ranges. Note that the wavenumber axis is shown here on a linear scale. The four panels show different projections of the CDM subhalo populations. The blue squares show the substructure power spectrum including all subhalos with masses less than $10^{10}$ M$_\odot$, while the other point types show the contributions from separate mass bins. We note that the contribution from the most massive subhalos included here ($10^9 M_\odot < M_{\rm sub}<10^{10}M_\odot$) varies significantly between different projections, with them making no contribution in the lower right panel.}\label{fig:mass_decomp}
\end{figure*}

In Figure \ref{fig:cat} we show how the power spectrum shape and amplitude change as a function of redshift (top) and highest subhalo mass included (bottom) for the CDM (left) and ETHOS4 (right) simulations. The fiducial cases are kept in the same color as in Figure \ref{fig:cdm_ethos4} (but notice that with $L=100$ kpc the two-subhalo term is no longer clearly discernible in the CDM simulation). For the redshit evolution, we consider three different epochs: $z = \{0,0.5,1\}$. These redshifts correspond to the redshift of the simulation snapshot from which the subhalo catalog was obtained.  For all cases, the source is assumed to be at $z=1.5$.

Since the convergence and the Einstein radius become ill-defined quantities as $z_{\rm lens} \rightarrow 0$, we artificially put our simulated $z=0$ lens galaxy at a redshift $z_{\rm lens}=0.5$ in order to compute their convergence field. In order words, we use the critical density for lensing $\Sigma_{\rm crit}$ corresponding to having a lens at $z=0.5$ and source at $z=1.5$ to compute the substructure convergence field of our simulated $z=0$ galactic halo. For our two other epochs ($z=0.5, 1$), $\Sigma_{\rm crit}$ is computed self-consistently using the redshift of the simulated halo as the lens redshift. It is therefore important to keep in mind that in the upper panels of Figure \ref{fig:cat}, the value of the critical density is changing between the $z=1$ and $z=0.5$ curves (by about at factor $\sim2$), but is the same for $z=0$ and $z=0.5$. This means that the relative amplitude between the $z=0$ and $z=0.5$ curves is really telling us something about subhalo accretion and evolution within the lens halo. 

The redshift dependence shown in Figure \ref{fig:cat} qualitatively agrees with what one would expect within the standard cosmological evolution: as we approach $z=0$, more subhalos are accreted into the host halos, implying that the amplitude of the power spectrum increases. This increase is more pronounced in the CDM case as more subhalos with $m < 10^8$ M$_\odot$ are accreted between $z=0.5$ and $z=0$ in this model. Also, as subhalos are accreted and move closer to the host center, mass loss due to tidal interaction becomes important. For the ETHOS4 simulation, we find that this leads on average to a reduction of the effective subhalo mass $m_{\rm eff}$ between $z=0.5$ and $z=0$, which partially compensates the slight increase in $\bar{\kappa}_{\rm sub}$ to leave the low-$k$ amplitude nearly unchanged\footnote{See the tables in Appendix \ref{sect:feat_tables} for the $m_{\rm eff}$ and $\bar{\kappa}_{\rm sub}$ values at the different redshifts.}. Furthermore, the much larger total number of subhalos in the CDM case also means that the two-subhalo term makes a non-negligible contribution at $z=0$, which tends to increase the magnitude of the redshift evolution in this case. In contrast, the ETHOS4 model does not get a significant two-subhalo contribution at $z=0$. 

Another important aspect of the redshift evolution is the difference in the power spectrum slope for $k\gtrsim 2$ kpc$^{-1}$. Again, this effect is more apparent in the ETHOS4 simulation than in the CDM simulation due to the lower central densities of subhalos in the former, making them more susceptible to tidal effects and mass loss. We indeed find that the substructure mass fraction in subhalos with $m < 10^7$ M$_\odot$ grows more rapidly between $z=0.5$ and $z=0$ in the ETHOS4 case compared to the CDM case, hence leading to a net transfer of power from larger to smaller scales in the power spectrum. This in turn results in a shallower slope for $k\gtrsim 2$ kpc$^{-1}$ at $z=0$ as compared to $z=0.5$.

The three different upper mass thresholds we consider in the lower panels of Figure \ref{fig:cat} are: $m_{\rm high} = \{10^8 \; \text{M}_{\odot},10^9 \; \text{M}_{\odot}\}$ and ``All subhalos'', where ``All subhalos'' means we include all subhalos above the resolution threshold. The behavior as a function of mass similarly shows the intuitive notion that, as we increase $m_{\rm high}$, the amplitude increases due to the fact that both $m_{\rm eff}$ and $N_{\rm sub}$ increase. The error bars are much larger for the case where all the subhalos are included because there are only a handful of subhalos with mass $> 10^9$ M$_{\odot}$, and they do not always get projected into the center-most region of the host. In the projections where even a single one of these subhalos is projected into the region of interest the amplitude is higher by about an order of magnitude. This shows that the low-$k$ amplitude ($k\lesssim1$ kpc$^{-1}$) is largely determined by the largest subhalos, as described in Refs.~\cite{Hezaveh_2014,Rivero:2017mao,Brennan:2018jhq}. Note that for the ``All subhalos'' ETHOS4 power spectrum the lower bound is very small. This is simply due to the fact that, except for the rare cases when a very massive subhalo gets projected into the center-most region, the highest subhalo mass across projections is nearly constant for this simulation. On the other hand, for the CDM simulation the upper mass bound displays more variation, which is why the lower bound is larger.

A question that often comes to mind when discussing the substructure power spectrum is which mass scale is this observable most sensitive to. It is generally assumed that the largest subhalos within the strong lensing region dominate the observable power spectrum signal, since subhalos of higher mass generally warp images more. However, what we find here is more subtle. Figure \ref{fig:mass_decomp} shows the decomposition of the dimensionless\footnote{The dimensionless power spectrum is defined (in 2 dimensions) as usual: $$\Delta^2_{\rm sub}(k) \equiv \frac{k^2P_{\rm sub}(k)}{2 \pi}.$$} convergence power spectrum into four different mass bins together with the power spectrum including all subhalos with masses below $10^{10}$ M$_{\odot}$ for four different projections in the CDM simulation (one in each panel). Surprisingly, it can be seen that the $10^7 - 10^8$ M$_{\odot}$ subhalos dominate the signal almost entirely on scales 2 kpc$^{-1} \lesssim k \lesssim 15$ kpc$^{-1}$. Subhalos with masses between $10^9 - 10^{10}$ M$_{\odot}$ are quite rare, and in fact sometimes are not even present (e.g.~lower right panel) in the strong lensing region. When present, they can of course dominate the signal at the lowest wavenumbers as discussed above, but they generally do not make the largest contribution to power spectrum on all observable scales. Another somewhat surprising element shown in Figure \ref{fig:mass_decomp} is the relatively small contribution that the $10^8 - 10^9$ M$_{\odot}$ subhalos make to the overall power spectrum. Despite being quite numerous and fairly massive, they have a lesser contribution to the overall signal than their less massive counterparts, except possibly at the lowest wavenumbers. 

\begin{figure*}[t!]
\centering
\includegraphics[width=0.47\textwidth]{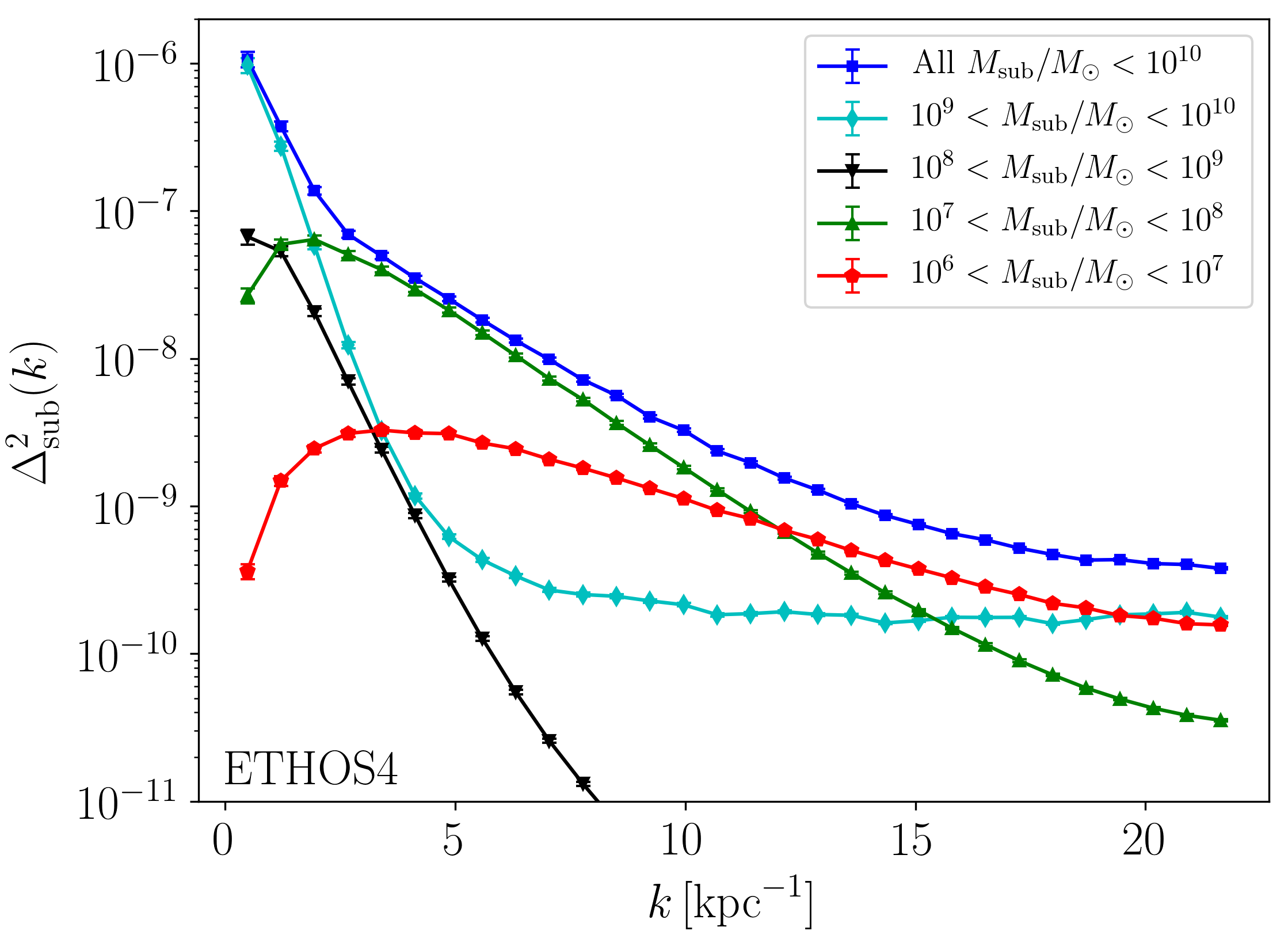}
\includegraphics[width=0.47\textwidth]{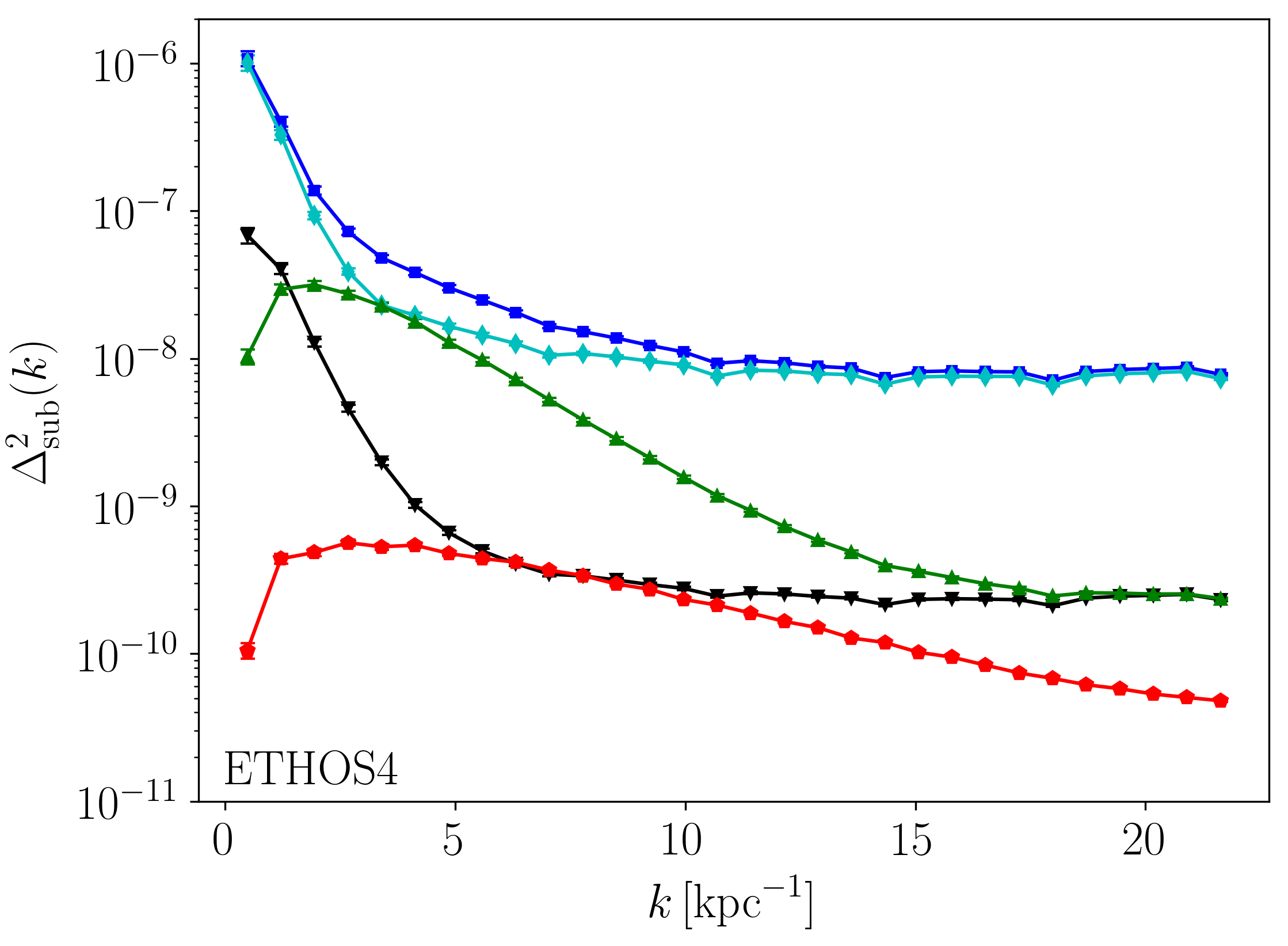}\\
\includegraphics[width=0.47\textwidth]{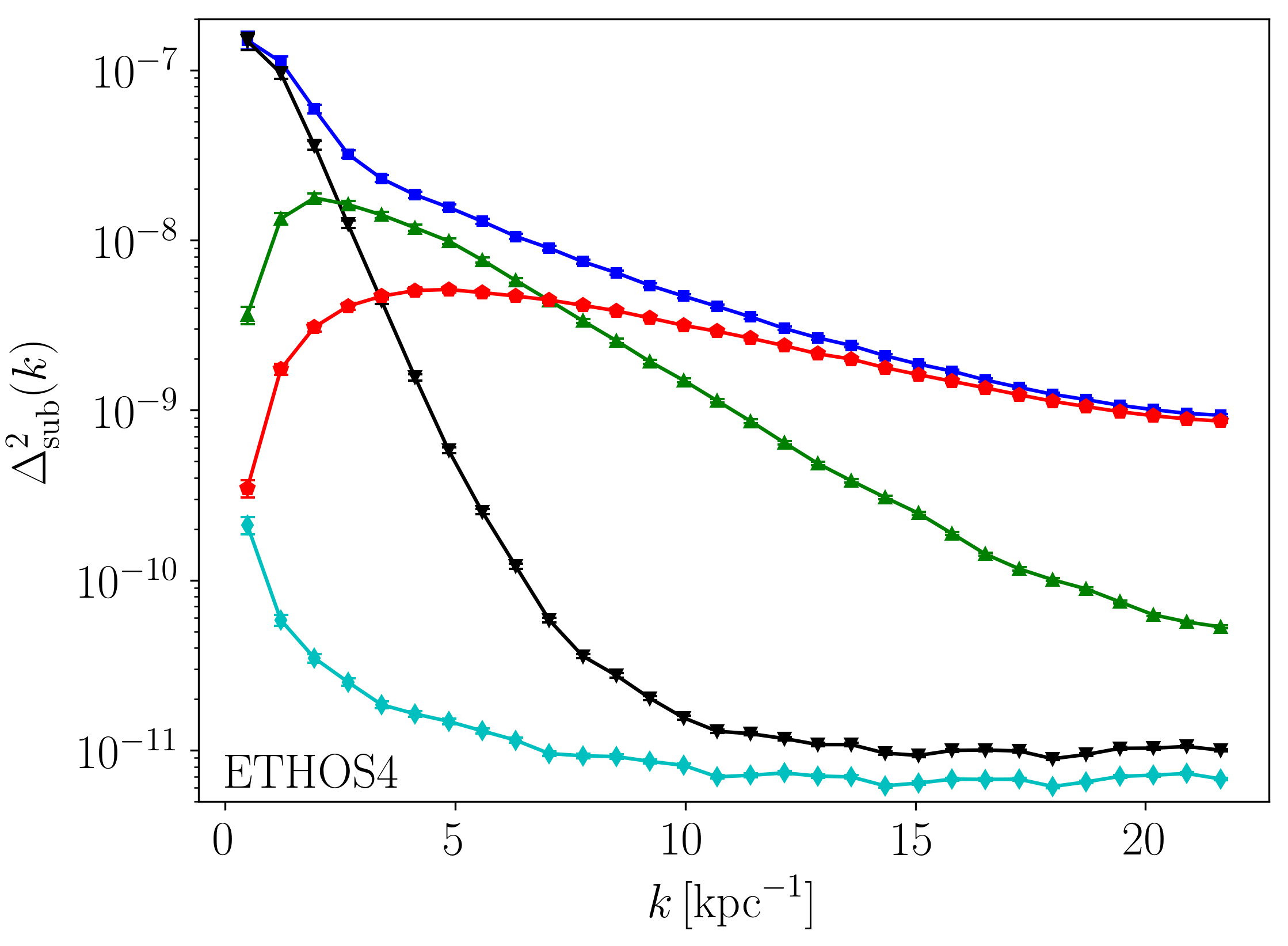}
\includegraphics[width=0.47\textwidth]{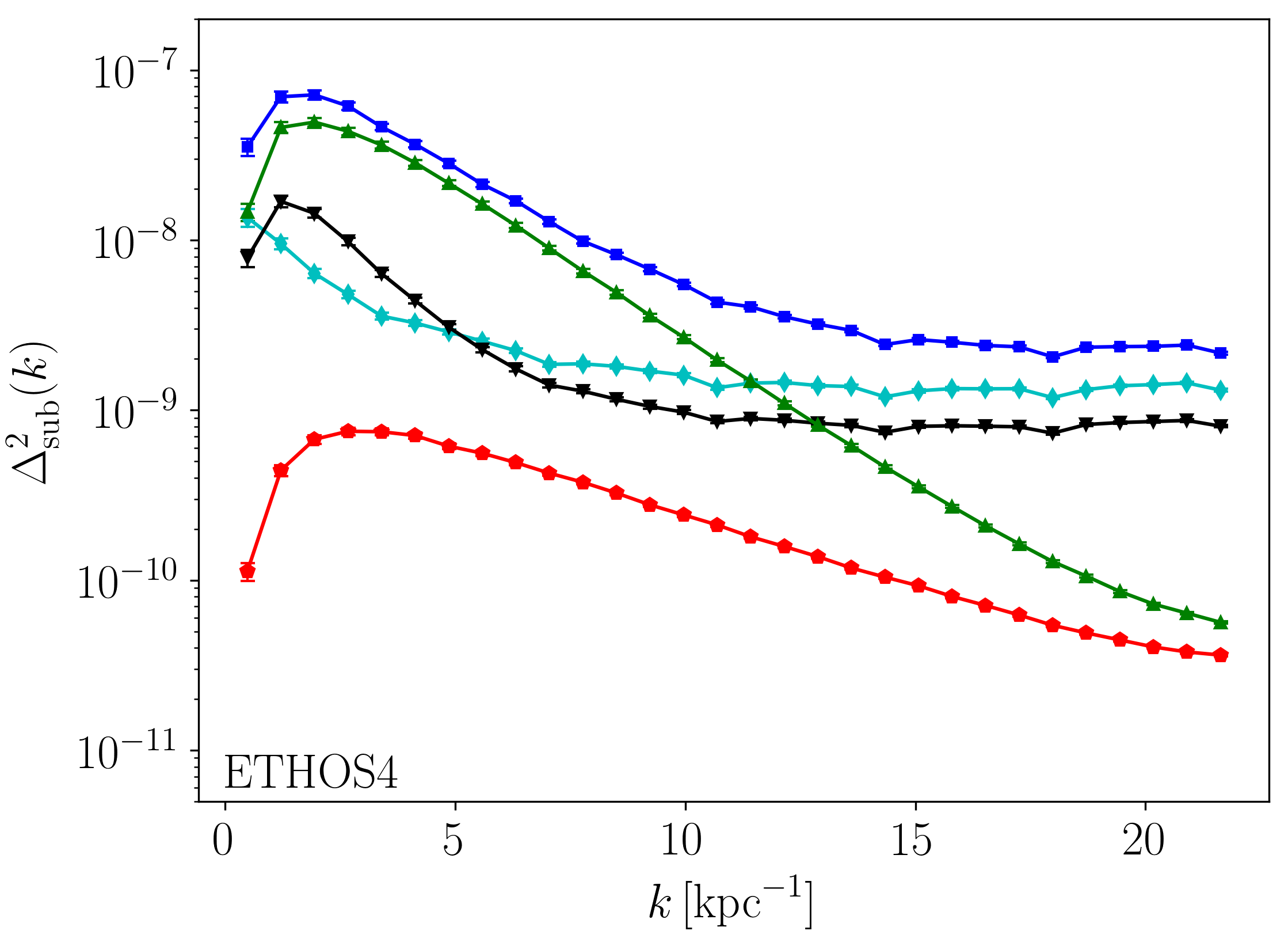}
\caption{Same as Figure~\ref{fig:mass_decomp} but for the ETHOS4 simulation.}\label{fig:mass_decomp_ethos4}
\end{figure*}

A similar decomposition is done for the ETHOS4 simulations, and it is shown in Figure \ref{fig:mass_decomp_ethos4}. It can be seen that the ETHOS4 projections display more variability than their CDM counterparts, due to the fact that there are many fewer subhalos. Even in this case, subhalos with mass $m<10^8$ M$_\odot$ seem to make on average a sizable contribution on scales 2 kpc$^{-1} \lesssim k \lesssim 15$ kpc$^{-1}$.

\bigskip 

\subsection{Snapshot particle data}\label{sec:results_part}

\begin{figure}[t!]
\centering
\includegraphics[width=0.5\textwidth]{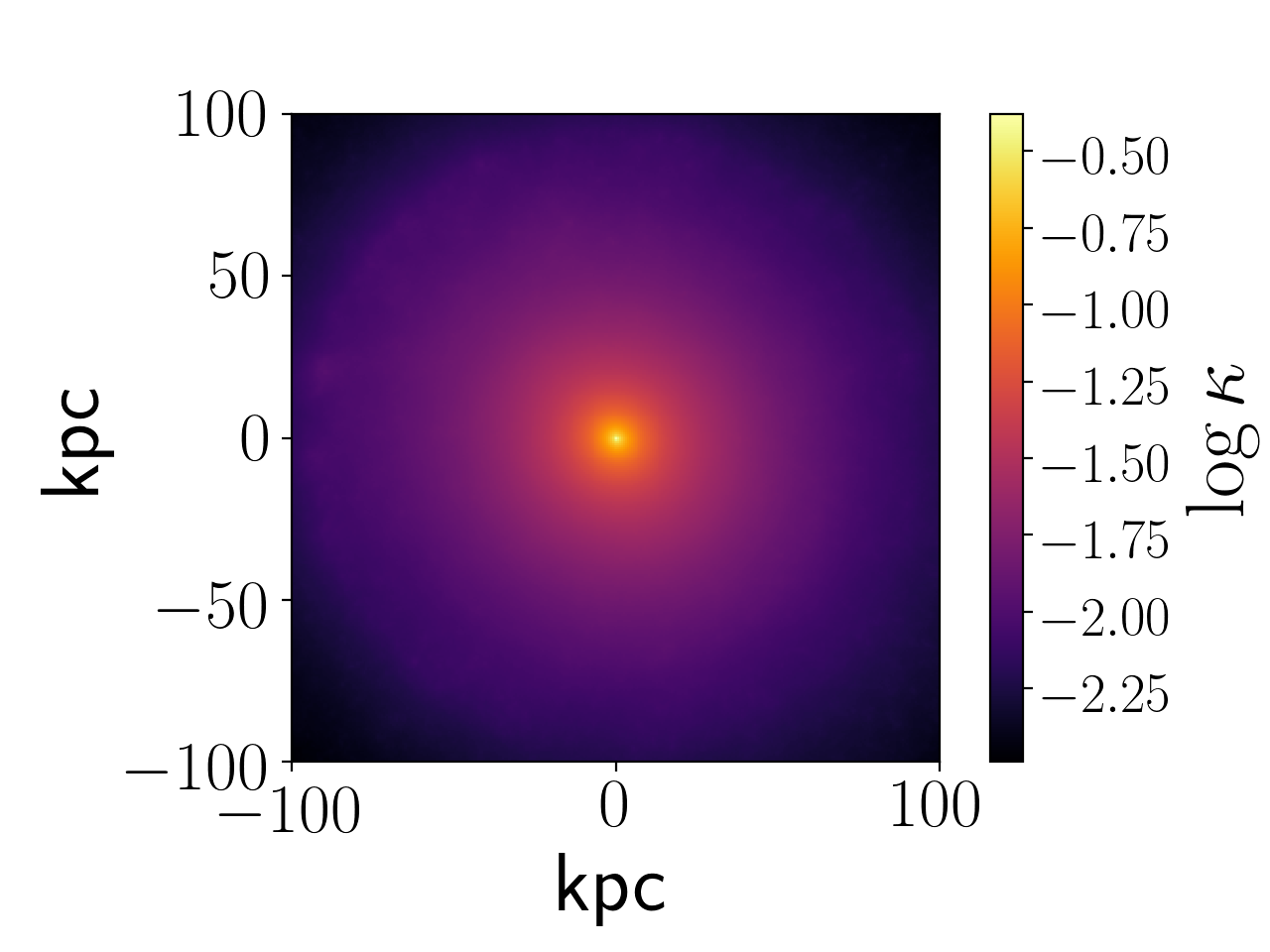}
\caption{Convergence field of the host in the CDM simulation found by averaging many projections along different LOS, as per Eq. \eqref{eq:kappa_host}.}\label{fig:host}
\end{figure}

\begin{figure*}[t!]
\centering
\begin{subfigure}
\centering
\includegraphics[width=0.8\textwidth]{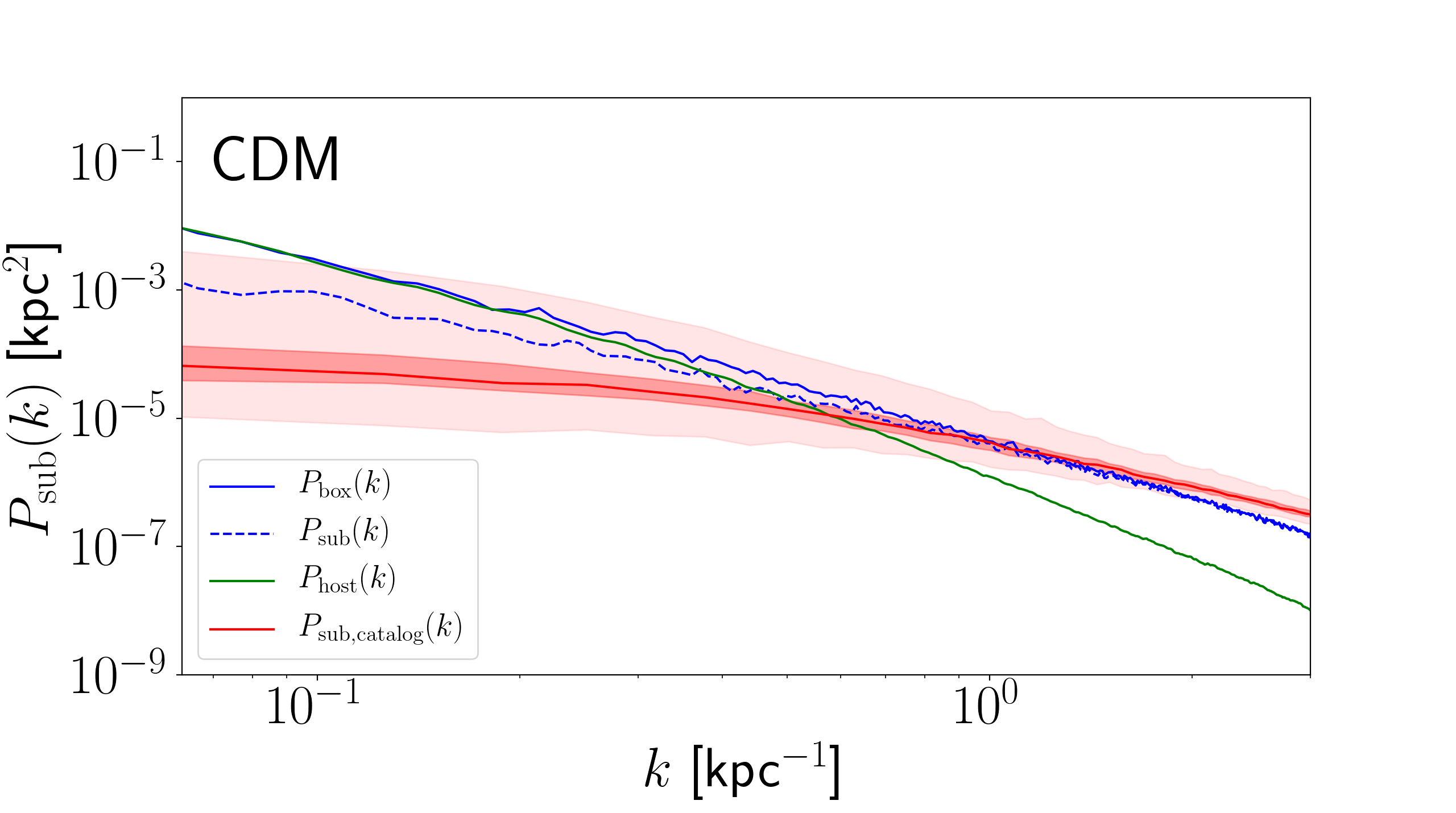}
\end{subfigure}
\begin{subfigure}
\centering
\includegraphics[width=0.8\textwidth]{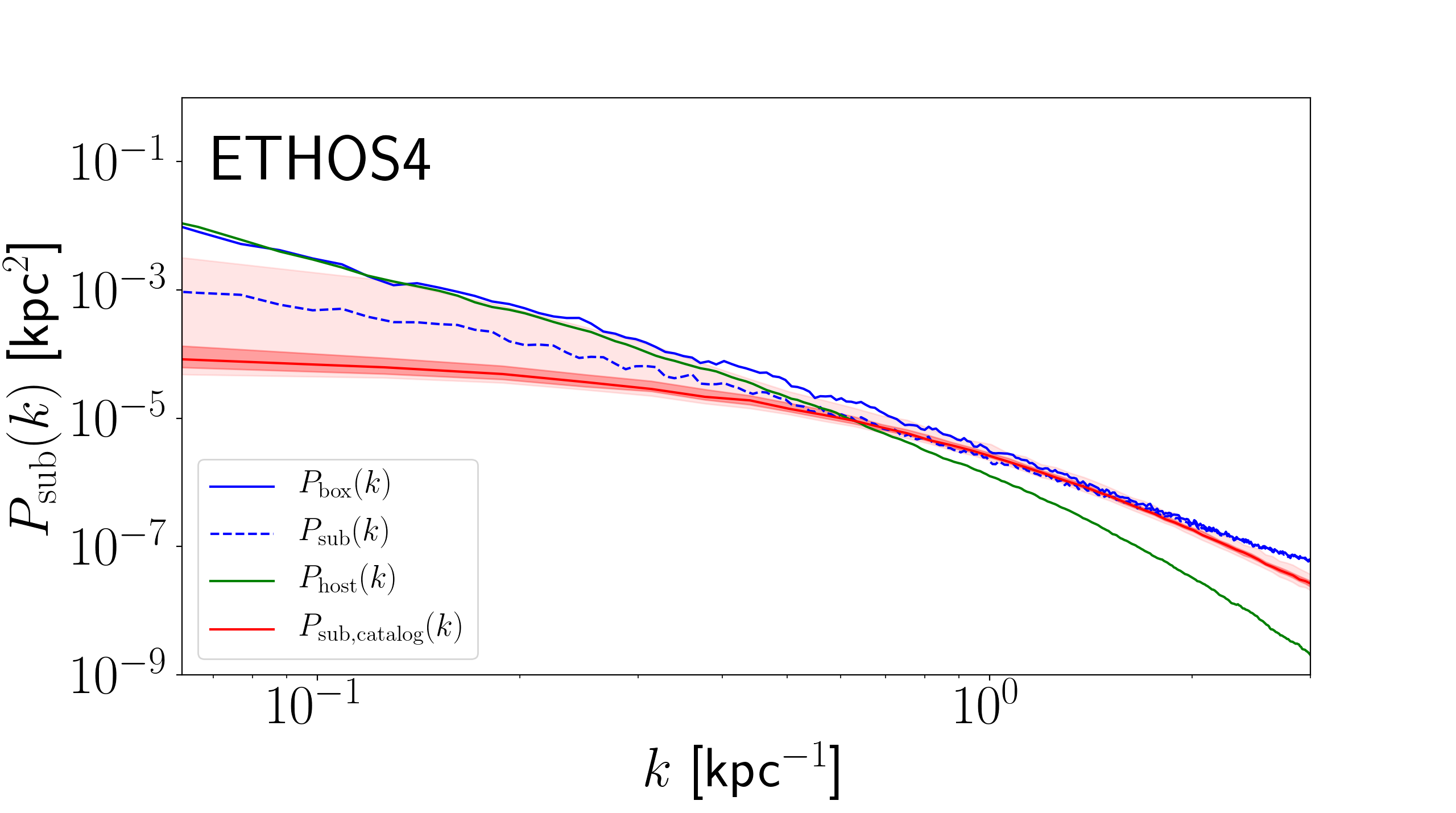}
\end{subfigure}
\caption{\label{fig:4} Power spectrum of the full simulation snapshots at $z=0.5$. The CDM (ETHOS4) simulation is in the top (bottom). The solid blue line is the power spectrum of the full projected field. The green line is the power spectrum of many projections averaged together, which approximates that of the host halo (as per Eq.~\eqref{eq:kappa_host}). The dashed blue line is the power spectrum of a projection with the average map subtracted, yielding the power spectrum due to substructure (as per Eq. \eqref{eq:pk_box-host}). In each plot we have overlaid the substructure power spectrum obtained from the catalogs in red, when all the subhalos are included (i.e. the same red lines as in the bottom two panels of Figure \ref{fig:cat}).}\label{fig:ps_particle}
\end{figure*}

The host convergence field obtained by averaging projections along different LOS in the CDM simulation is shown in Figure \ref{fig:host}. 
The power spectra for both simulation suites are displayed in Figure \ref{fig:ps_particle}. The top (bottom) panel corresponds to the CDM (ETHOS4) simulation. The solid blue line is the power spectrum obtained from a single projection of the $N$-body particles, without having performed any host subtraction. The dashed blue line is the power spectrum after removing the host contribution from a projection map, thus approximating the power spectrum due to the substructure, as per Eq.~\eqref{eq:pk_box-host}. The green line is that of the average convergence map, i.e.~approximately the host (the Fourier transform of Figure \ref{fig:host}). 

There are several notable features in these figures. First of all, notice the suppression in power at high $k$ of the green lines compared to the solid blue lines, which shows that the averaging procedure is indeed removing the contribution from substructure on these scales. Furthermore, when the host is subtracted (dashed lines), a lot of power is lost at low $k$ but conversely we regain the power on small scales, which corresponds to the substructure convergence field remaining after the host is removed. 

The overall amplitude is the same for both simulations, since both simulations have roughly the same number of particles in the region of interest. However, in the host and in the substructure power spectra at high $k$ (>1 kpc$^{-1}$) we can still see the suppression of power of ETHOS4 with respect to CDM due to the cutoff in the initial matter power spectrum (and to the self-interactions, albeit to a lesser extent), which causes both a suppressed number of small-mass subhalos and reduced central densities for the remaining ones.

In both figures we also overlay the catalog power spectra with no high mass cut (i.e. the same two red lines as in Figure \ref{fig:cat}). Unexpectedly, at high $k$ the amplitude of the power spectra derived from the catalogs is higher than that obtained from the corresponding particle snapshot when looking at the CDM simulation (and comparable in the ETHOS4 simulation). We expected the opposite since when we Fourier transform the full simulation box we are capturing all the substructure (e.g.tidal debris), not just objects found by the halo finder. However this can be understood by considering the discrete nature of the simulation particles, and the fact that when we reach very small scales (i.e. around the scale radius of subhalos) there are in fact only a handful of particles. By instead imposing a smooth, truncated NFW profile at the catalog level we are artificially boosting the high-$k$ signal with respect to the particle power spectrum. This effect is not quite as palpable in the ETHOS4 simulation since the truncated Burkert fit is cored in the central regions. By looking at Figure \ref{fig:projections} and comparing the projections from the simulation snapshots and those built from the subhalo catalogs it becomes apparent that a part of this discrepancy might also be due to the loss of ellipticity when imposing spherically-symmetric convergence profiles. 

Finally, notice that at small $k$ the dashed lines lie within the 90\% confidence band of the catalog power spectrum. This is indicative of the fact that the very large substructure in the lens is well captured by the halo finder, and since said structure dominates the amplitude (as shown in the previous section), the particle and catalog power spectra are similar on these scales. 

\section{Discussion and Conclusion}\label{sec:discussion}

In this paper we have provided a comprehensive study of the substructure convergence power spectrum in $N$-body simulations. By comparing this observable in two simulations within the ETHOS framework ~\cite{Vogelsberger:2015gpr} that differ in their treatment of dark matter microphysics (both at early and late times through dark matter-dark radiation interactions in the former and DM-DM self-interactions in the latter) we have been able to identify different ways in which details in the particle nature of dark matter can come to light through this observable. 

We chose to carry out our analysis in two different, but complimentary, ways. On one hand, we have an idealized scenario in which all substructure is perfectly spherical and identifiable, and can be fit with simple density profiles. On the other hand, we have a scenario in which there are no assumptions or definitions built into what is considered to be a subhalo - instead we capture all the structure within the host halo. The former method of course benefits from its simplicity: it allows us to clearly disentangle different subhalo properties and their impact on the shape and amplitude of the power spectrum. The latter, however, more closely approximates reality, where one cannot choose what perturbs an image or an arc, and has to carefully think about how to account for the mass distribution of the main lens galaxy itself. 

Doing the catalog analysis we confirmed many of the properties outlined by Ref.~\cite{Rivero:2017mao} and brought to light several new ones. We were able to show how the amplitude and shape of the power spectrum are related to the abundance, sizes, and masses of subhalos. Furthermore, we showed the redshift evolution of the power spectrum, and saw a difference in the standard CDM vs.~DM-DR+SIDM scenarios: in the former we observed an expected increase in the amplitude of the power spectrum as more substructure was accreted; conversely, in the latter, between $z=0$ and $z=0.5$ there was nearly no change in the amplitude in these two redshift bins. This was partly the result of the lower total number of subhalos accreted during that time span in ETHOS4, as well the higher susceptibility of ETHOS4 subhalos to tidal disruption which caused $m_{\rm eff}$ to shrink within the strong lensing region. A non-negligible two-subhalo contribution at $z=0$ for CDM also helps explain the faster growth of the overall power spectrum amplitude in this latter case.

The other interesting effect that came to light when comparing the ETHOS4 redshift evolution between $z=0$ and $z=0.5$ was the difference in slope on scales $k \gtrsim 2$ kpc$^{-1}$, which reflected the changing subhalo mass function as the host evolves. Both of these effects that appear in the redshift evolution of ETHOS4 - the amplitude and the slope - offer exciting possibilities. In Ref.~\cite{Rivero:2017mao} the highest $k$ values ($\gtrsim 100$ kpc$^{-1}$) were identified as the most interesting region in the power spectrum to study the particle nature of dark matter. This was unfortunate since it is unlikely that we will be able to measure modes past $k \sim 100$ kpc$^{-1}$ in the near future, and baryonic structures of comparable sizes would interfere with the isolation of the dark matter power spectrum slope on these scales. However, here we have identified other ways of probing dark matter microphysics that involve scales that can in fact be probed with current and future observations ($0.1 \leq k/$kpc$^{-1} < 100$).
 
Our results highlight the fact that combining different lenses in order to boost the signal-to-noise of a substructure power spectrum measurement is highly non-trivial. Indeed, detailed models for the redshift evolution of the subhalo population for different host properties would have to be included in the fit. The other side of that coin is that when (if) observations are good enough to measure the power spectrum with a single lens, comparing the low-$k$ amplitude with lenses at different redshifts can serve as a diagnostic tool for dark matter deviations from standard CDM, be it an effect at early time that is imprinted on the initial power spectrum and consequently delays structure formation, or an effect at late times, like self-interactions that are strong (or weak but inelastic \cite{Vogelsberger:2018bok}) enough to cause subhalo stripping and/or disruption. One would of course also have to consider how the presence of baryons can disrupt substructure as well.

Furthermore, by looking into the mass decomposition of the power spectrum we found that it is not exclusively sensitive to the most massive subhalos. For instance, we found that the mass range $10^7-10^8$ M$_{\odot}$ tends to dominate the power spectrum on intermediate scales (2 kpc$^{-1} \lesssim k \lesssim 15$ kpc$^{-1}$), particularly in the CDM simulation. In more standard gravitational imaging searches for substructure, sensitivity is assumed to be an increasing function of mass (and proximity to the images/arcs; see also e.g. Ref. \cite{Vegetti:2014wza} to see how other parameters, like concentration, can affect distortions). In this different, statistical approach we can see that this is no longer necessarily the case, and observations could probe lower masses in the highly coveted subhalo mass function. 

An important point to keep in mind is that the MW-like halos we have considered in this work are not typical lens galaxies (recall that both halos are sub-critical, i.e. $\kappa < 1$). Gravitational lenses at cosmological distances from our own galaxy in general have to be more massive and dense to act as strong lenses (see, e.g., Ref.~\cite{Bolton2008}). Such galaxies are expected to have more substructure (since substructure content scales with host mass), so the results presented in this paper should be considered as a useful lower limit on the amplitude of the convergence power spectrum in strong lens galaxies. One should also keep in mind that the increased central density in more typical strong lenses could increase subhalo tidal disruption \cite{Fiacconi:2016cih}. Quantifying these effects as a function of host halo density is an interesting future step in understanding this observable. 

By comparing the power spectrum obtained directly from the $N$-body particles with that from the subhalo catalogs we actually saw that the halo model-based computation (as used in Ref.~\cite{Rivero:2017mao}) is in fact an excellent approximation to the more detailed density field (as captured by the simulation snapshots). As we mentioned above, this does break down at high $k$, where the imposition of a smooth convergence profile leads to an overestimation with respect to the $N$-body particle power spectrum on those same scales due to the finite spatial resolution of the mass particles. On intermediate scales, the difference between the catalog and particle power spectra is well within forecasted error bars for the convergence power spectrum \cite{Cyr-Racine:2018htu}. This result lends weight to the robustness of this observable to study substructure populations at cosmological distances from the Milky Way.

A crucial extension to this work is going to involve carrying this analysis out with a hydrodynamical simulation in order to make robust predictions that can be compared with observations in the near future, since we know that on these scales baryonic processes can have quite significant effects on the dark matter distribution. Some work has been carried out to study the difference in distortions due to a population of globular clusters versus dark matter subhalos \cite{He:2017qdo}, showing that milliarcsecond resolution images could distinguish between these in direct detection efforts. But the impact on the power spectrum has yet to be addressed. 

As mentioned above, it will be particularly interesting to study the redshift evolution of the amplitude and the slope on observable scales in these simulations, since this can be a key way of distinguishing CDM from alternative dark matter scenarios that cause substructure disruption. Another aspect that we leave for future work is considering the placement of the high resolution box within the environment of the full cosmological simulation. Recently there has been a lot of interest on the relative importance of perturbations to lensed images by substructure intrinsic to the lens versus along the line of sight \cite{Keeton:2003aa,Li:2016afu,Despali:2017ksx,Birrer:2016xku}. Indeed it is crucial to understand this well if we want to be able to falsify or confirm CDM with this observable. If all the identified perturbers to a given image are assumed to be virially bound to the host one can overestimate the fraction of substructure in the host, and thus make an erroneous comparison to a standard CDM prediction. Thus, quantifying the contribution from LOS substructure is crucial in advancing the capability of strong lensing to constrain the particle nature of dark matter.

%%%%%%%%%%%%%%%%%%%%%%%
\acknowledgments
CD and ADR were supported by NSF grant AST-1813694, Department of Energy (DOE) grant DE-SC0019018, and the Dean's Competitive Fund for Promising Scholarship at Harvard University. F.-Y. C.-R.~acknowledges the support of the National Aeronautical and Space Administration (NASA) ATP grant NNX16AI12G at Harvard University. MV acknowledges support through an MIT RSC award, the Alfred P. Sloan Foundation, NASA ATP
grant NNX17AG29G, and a Kavli Research Investment Fund. JZ  acknowledges  support  by  a Grant of Excellence from the 
Icelandic Research Fund (grant number 173929-051).

\appendix

\section{Truncated convergence profiles} \label{sect:conv_profiles}

The truncated NFW profile (tNFW) \cite{Baltz:2007vq} is given by

\begin{equation}\label{eq:3DNFW}
\rho_{\rm tNFW}(r) = \frac{m_{\textsc{\tiny NFW}}}{4\pi r (r+r_{\rm s})^2}\left(\frac{r_{\rm t}^2}{r^2+r_{\rm t}^2}\right),
\end{equation}
where $r_{\rm s}$ is the scale radius, $r_{\rm t}$ is the tidal radius and $m_{\textsc{\tiny NFW}}$ is defined below. Integrating this profile along the line of sight and diving by the critical density for lensing $\Sigma_{\rm crit}$, we obtain the tNFW convergence profile:

\begin{align}\label{eq:smd_tnfw}
\kappa_{\rm tNFW}(x) &= \frac{m_{\textsc{\tiny NFW}}}{ \Sigma_{\rm crit}r_{\rm s}^2} \frac{\tau^2}{2 \pi  (\tau^2 + 1)^2} \Bigg[\frac{\tau^2+1}{x^2-1}(1-F(x))   \nonumber\\
& + 2F(x) - \frac{\pi}{\sqrt{\tau^2 + x^2}} + \frac{\tau^2-1}{\tau \sqrt{\tau^2 + x^2}} L(x) \Bigg],
\end{align}
where
\begin{align}\label{eq:x_tau}
x=\frac{r}{r_{\rm s}},\quad \tau = \frac{r_{\rm t}}{r_{\rm s}},
\end{align}
\begin{align}
F(x) = \frac{\cos^{-1}(1/x)}{\sqrt{x^2-1}},
\end{align}
\begin{align}
L(x) = \ln\left(\frac{x}{\sqrt{\tau^2 + x^2} + \tau} \right),
\end{align}
and 
\begin{align}
m = \frac{m_{\textsc{\tiny NFW}} \tau^2}{(\tau^2+1)^2}\left[(\tau^2-1)\ln(\tau) + \tau \pi - (\tau^2 + 1)\right].
\end{align}

The truncated Burkert profile \cite{Burkert:1995yz,Rivero:2017mao} is given by:
\begin{equation}\label{eq:modified_burkert}
\rho_{\rm tBurk}(r) = \frac{m_{\rm b}}{4 \pi (r + p \; r_{\rm s})(r^2 + p^2 r_{\rm s}^2)} \left(\frac{r_{\rm t}^2}{r^2 + r_{\rm t}^2}\right),
\end{equation}
where $r_{\rm b}$ is the core radius, and the scale mass $m_{\rm b}$ is the mass within the core.  Here we set $r_{\rm b} = p \,r_{\rm s} $, where $p$ is a constant that represents the size of the core as a fraction of the scale radius. The convergence field of this density profile is then \cite{Rivero:2017mao}:

\begin{align}\label{eq:tburk_conv}
\kappa_{\rm tBurk}(x) &= \frac{m_{\rm b}}{8 \pi \Sigma_{\rm crit} r_{\rm s}^2} \; \tau^2 \Bigg\{ \pi \Bigg( \frac{2p \sqrt{\frac{1}{\tau^2 + x^2}}}{p^4 - \tau^4} - \frac{\sqrt{\frac{1}{x^2-p^2}}}{p(\tau^2 + p^2)} \nonumber \\
& - \frac{\sqrt{\frac{1}{x^2 +p^2}}}{p^3 -p\tau^2} \Bigg) + \frac{2 \arctan \left[\frac{p}{\sqrt{x^2 - p^2}} \right]}{\sqrt{x^2 - p^2}(p^3 + p\tau^2)}- \\
& \frac{2  \tanh^{-1} \left[\frac{p}{\sqrt{p^2 + x^2}} \right]}{\sqrt{x^2 +p^2}(p^3-p \tau^2)} + \frac{4 \tau \tanh^{-1} \left[\frac{\tau}{\sqrt{x^2 +\tau^2}} \right]}{\sqrt{x^2 + \tau^2}(p^4-\tau^4)} \Bigg\},\nonumber
\end{align}
where again $x$ and $\tau$ are defined as in Eq. \eqref{eq:x_tau}.

\begin{widetext}

\section{Features of the convergence maps} \label{sect:feat_tables}

The tables in this section display some quantities of interest extracted from the two simulations. The main value quoted corresponds to the median across 90 projections for a box size with $L=100$ kpc, while the errors correspond to the 90\% confidence interval. The exception to this is the first entry in each table, $N_{\rm sub}(L=300$ kpc), since this is a quantity extracted from the original simulations before doing any projections. The Einstein radius is fixed to 1 arcsecond. With our cosmology, this corresponds to $R_{\rm Ein} = 6.18$ kpc at $z=0.5$\footnote{We will be using this value in the $z=0$ tables as well for the reasons outlined in \ref{sec:results_cat}.} and $R_{\rm Ein} = 8.10$ kpc at $z=1$.

Notice that in the "All subhalos" column the $m_{\rm eff} \equiv \langle m^2\rangle / \langle m \rangle$ entry displays very large upper bounds. This is due to the fact that when there are very few subhalos with masses $> 10^9$ M$_{\odot}$ they only rarely get projected into the region of interest. Even a one order of magnitude difference in the maximum subhalo mass can change $m_{\rm eff}$ considerably, yielding a very large upper bound. This is reflected in the large 90\% confidence interval in the red lines of Figure \ref{fig:cat}.

\subsection{CDM}

\subsubsection{z=0}

\begin{minipage}{\linewidth}
  \footnotesize
    \bigskip
 \centering
 \begin{tabular}{ | c | c | c | c |}
\hline
 & $m_{\rm high}=10^8$ M$_{\odot}$& $m_{\rm high} = 10^9$ M$_{\odot}$  & All subhalos   \\ \hline
 $N_{\rm sub}$($L_{\rm box}=300$ kpc)  & 9810   & 10007 & 10031 \\ [0.2cm]
 $N_{\rm sub}$($L_{\rm box} = 100$ kpc) & $1004^{+116}_{-97}$ & $1026^{+148}_{-116}$ & $1024^{+104}_{-109}$  \\ [0.2cm]
 $N_{\rm sub}(R_{\rm Ein}$)  & $12^{+8}_{-6}$   & $13^{+8}_{-5}$   & $13^{+7}_{-6}$                   \\ [0.2cm]
 $\bar{\kappa}_{\rm sub}$ & $\left(3.28^{+0.54}_{-0.37}\right) \times 10^{-4} $ & $\left(5.06 ^{+1.19}_{-0.97} \right) \times 10^{-4} $ & $\left(1.22 ^{+0.94}_{-0.51} \right) \times 10^{-3}$ \\ [0.2cm]
 $\langle m \rangle$ [M$_{\odot}$]  & $\left(7.75^{+0.41}_{-0.57}\right)\times 10^{6}$ & $\left(1.20^{+0.18}_{-0.20}\right)\times 10^7$ & $\left(2.99^{+2.34}_{-1.33}\right)\times 10^7$ \\ [0.2cm]
 $m_{\rm eff} \equiv \langle m^2 \rangle / \langle m \rangle$ [M$_{\odot}]$ & $\left(2.72^{+0.17}_{-0.27}\right)\times 10^7$ & $\left(1.64^{+0.83}_{-0.83}\right) \times 10^8$ & $\left(7.79^{+20.1}_{-7.30}\right) \times 10^9$ \\ [0.2cm]
 $r_{\rm t,max}$ [kpc]  & $8.17^{+11.67}_{-2.70}$ & $9.51^{+5.53}_{-3.33}$ & $10.96^{+26.50}_{-2.88}$           \\ [0.2cm]
 $r_{\rm s,min}$ [kpc] & $0.03^{+0.00}_{-0.01}$ & $0.03^{+0.00}_{-0.01}$ & $0.03^{+0.00}_{-0.01}$             \\ [0.2cm]
\hline
\end{tabular}\par
Table A1.1
\end{minipage}

\subsubsection{z=0.5}

\begin{minipage}{\linewidth}
  \footnotesize
    \bigskip
 \centering
 \begin{tabular}{ | c | c | c | c |}
\hline
   & $m_{\rm high}=10^8$ M$_{\odot}$       & $m_{\rm high} = 10^9$ M$_{\odot}$     & All subhalos \\ \hline
 $N_{\rm sub}(L_{\rm box}=300$ kpc)   & 6516 & 6651 & 6669  \\ [0.2cm]
 $N_{\rm sub}(L_{\rm box}=100$ kpc) & $587^{+70}_{-57}$  & $584^{+103}_{-59}$ & $596^{+88}_{-59}$ \\ [0.2cm]
 $N_{\rm sub}(R_{\rm Ein})$  & $9^{+3}_{-4}$ & $9^{+5}_{-5}$  & $9^{+5}_{-5}$ \\ [0.2cm]
 $\bar{\kappa}_{\rm sub}$  & $\left(1.90^{+0.16}_{-0.22}\right) \times 10^{-4} $ & $\left(2.93^{+0.72}_{-0.58}\right) \times 10^{-4}$ & $\left(4.18^{+7.87}_{-1.55}\right)\times 10^{-4}$ \\ [0.2cm]
 $\langle m \rangle$ [M$_{\odot}$]  & $\left(7.67^{+0.48}_{-0.58}\right) \times 10^6$ & $\left(1.20^{+0.23}_{-0.19}\right) \times 10^7$ & $\left(1.66^{+3.04}_{-0.58}\right)\times10^7$ \\ [0.2cm]
 $m_{\rm eff} \equiv \langle m^2 \rangle / \langle m \rangle$ [M$_{\odot}]$ & $\left(2.53^{+0.45}_{-0.37}\right) \times10^7$ & $\left(1.09^{+1.04}_{-0.44}\right)\times10^8$ & $\left(5.95^{+128.00}_{-4.99}\right) \times 10^8$ \\ [0.2cm]
 $r_{\rm t,max}$ [kpc]  & $9.62^{+4.14}_{-3.95}$ & $10.90^{+6.38}_{-2.94}$  & $12.19^{+6.12}_{-2.87}$\\ [0.2cm]
 $r_{\rm s,min}$ [kpc]  & $0.06^{+0.02}_{-0.02}$ & $0.07^{+0.02}_{-0.02}$  & $0.06^{+0.02}_{-0.02}$\\ [0.2cm]
\hline
\end{tabular}\par
Table A1.2
\end{minipage}

\subsubsection{z=1}

\begin{minipage}{\linewidth}
  \footnotesize
    \bigskip
 \centering
 \begin{tabular}{ | c | c | c | c |}
\hline
  & $m_{\rm high}=10^8$ M$_{\odot}$    & $m_{\rm high} = 10^9$ M$_{\odot}$  & All subhalos  \\ \hline
 $N_{\rm sub}$($L_{\rm box}=300$ kpc) & 4694  & 4783   & 4798  \\ [0.2cm]
 $N_{\rm sub}$($L_{\rm box} = 100$ kpc) & $446^{+80}_{-49}$ & $459^{+75}_{-46}$ & $453^{+90}_{-54}$ \\ [0.2cm]
 $N_{\rm sub}(R_{\rm Ein}$) & $11^{+4}_{-6}$ & $11^{+4}_{-6}$ & $11^{+5}_{-6}$ \\ [0.2cm]
 $\bar{\kappa}_{\rm sub}$  & $\left(7.93^{+1.71}_{-1.26} \right)\times 10^{-5}$ & $\left(1.52^{+0.53}_{-0.52}\right)\times 10^{-4}$ & $\left(1.74^{+2.63}_{-0.64} \right) \times 10^{-4}$ \\ [0.2cm]
 $\langle m \rangle$ [M$_{\odot}$] & $\left(7.57^{+0.66}_{-0.58}\right) \times 10^6$ & $\left(1.43^{+0.36}_{-0.36}\right)\times 10^7$ & $\left(1.66^{+1.91}_{-0.51}\right)\times 10^7$ \\ [0.2cm]
 $m_{\rm eff} \equiv \langle m^2 \rangle / \langle m \rangle$ [M$_{\odot}]$ & $\left(2.60^{+0.33}_{-0.38}\right)\times 10^7$ & $\left(2.05^{+1.14}_{-0.95}\right)\times 10^8$ & $\left(2.82^{+44.1}_{-1.64}\right) \times 10^8$ \\ [0.2cm]
 $r_{\rm t,max}$ [kpc] & $10.12^{+7.08}_{-3.28}$ & $14.27^{+6.15}_{-5.47}$ & $16.00^{+10.70}_{-5.72}$ \\ [0.2cm]
 $r_{\rm s,min}$ [kpc] & $0.08^{+0.03}_{-0.01}$ & $0.08^{+0.03}_{-0.02}$ & $0.08^{+0.03}_{-0.02}$ \\ [0.2cm]
\hline
\end{tabular}\par
Table A1.3
\end{minipage}

\subsection{ETHOS4}

\subsubsection{z=0}

\begin{minipage}{\linewidth}
  \footnotesize
    \bigskip
 \centering
 \begin{tabular}{ | c | c | c | c |}
\hline
 & $m_{\rm high}=10^8$ M$_{\odot}$    & $m_{\rm high} = 10^9$ M$_{\odot}$  & All subhalos  \\ \hline
 $N_{\rm sub}$($L_{\rm box}=300$ kpc)  & 821 & 898  & 918 \\ [0.2cm]
 $N_{\rm sub}$($L_{\rm box} = 100$ kpc) & $93^{+28}_{-15}$ & $100^{+35}_{-15}$  & $101^{+47}_{-14}$ \\ [0.2cm]
 $N_{\rm sub}(R_{\rm Ein}$)  & $1^{+2}_{-1}$ & $2^{+2}_{-2}$ & $2.00^{+2}_{-2}$ \\ [0.2cm]
 $\bar{\kappa}_{\rm sub}$  & $\left(4.06^{+1.33}_{-1.08}\right) \times 10^{-5}$ & $\left(1.28^{+0.49}_{-0.59}\right) \times 10^{-4}$ 
 & $\left(7.36^{+12.80}_{-5.65}\right)\times 10^{-4}$ \\ [0.2cm]
 $\langle m \rangle$ [M$_{\odot}$]  & $\left(1.05^{+0.15}_{-0.19}\right)\times 10^7$ & $\left(2.92^{+1.77}_{-1.11}\right) \times 10^7$ & $\left(1.89^{+2.58}_{-1.47}\right) \times 10^8$ \\ [0.2cm]
 $m_{\rm eff} \equiv \langle m^2 \rangle / \langle m \rangle$ [M$_{\odot}]$ & $\left(3.07^{+0.84}_{-0.90}\right) \times 10^7$ & $\left(3.63^{+1.79}_{-2.40}\right) \times 10^8$ & $\left(1.08^{+2.35}_{-1.02}\right) \times 10^{10}$ \\ [0.2cm]
 $r_{\rm t,max}$ [kpc]  & $10.86^{+18.13}_{-3.54}$  & $13.85^{+12.21}_{-4.80}$ & $20.28^{+45.90}_{-6.99}$ \\ [0.2cm]
 $r_{\rm s,min}$ [kpc] & $0.06^{+0.00}_{-0.02}$ & $0.06^{+0.00}_{-0.02}$ & $0.06^{+0.00}_{-0.02}$ \\ [0.2cm]
\hline
\end{tabular}\par
Table A2.1
\end{minipage}

\subsubsection{z=0.5}

\begin{minipage}{\linewidth}
  \footnotesize
    \bigskip
 \centering
 \begin{tabular}{ | c | c | c | c |}
\hline
  & $m_{\rm high}=10^8$ M$_{\odot}$  & $m_{\rm high} = 10^9$ M$_{\odot}$  & All subhalos   \\ \hline
 $N_{\rm sub}$($L_{\rm box}=300$ kpc)   & 579 & 629 & 642  \\ [0.2cm]
 $N_{\rm sub}$($L_{\rm box} = 100$ kpc)  & $57^{+19}_{-10}$ & $60^{+19}_{-8}$ & $62^{+17}_{-8}$ \\ [0.2cm]
 $N_{\rm sub}(R_{\rm Ein}$)  & $0.50^{+1.5}_{-0.5}$ & $0.5^{+2}_{-0.5}$ & $0.5^{+2}_{-0.5}$ \\ [0.2cm]
 $\bar{\kappa}_{\rm sub}$ & $\left(3.43^{+1.38}_{-1.02}\right)\times 10^{-5}$ & $\left(7.17^{+2.71}_{-3.30}\right)\times 10^{-5}$ & $\left(1.91^{+8.38}_{-0.77}\right)\times 10^{-4}$ \\ [0.2cm]
 $\langle m \rangle$ [M$_{\odot}$]  & $\left(1.45^{+0.38}_{-0.32}\right) \times 10^7$ & $\left(2.77^{+1.16}_{-1.01}\right)\times 10^7$ & $\left(8.06^{+37.30}_{-3.25}\right)\times 10^7$ \\ [0.2cm]
 $m_{\rm eff} \equiv \langle m^2 \rangle / \langle m \rangle$ [M$_{\odot}]$ & $\left(3.75^{+0.70}_{-0.76}\right)\times 10^7$ & $\left(1.47^{+3.26}_{-0.88}\right) \times 10^8$ & $\left(1.31^{+17.30}_{-0.36}\right)\times 10^9$ \\ [0.2cm]
 $r_{\rm t,max}$ [kpc]  & $13.33^{+59.01}_{-5.75}$  & $20.22^{+34.73}_{-10.95}$ & $26.57^{+25.22}_{-14.72}$ \\ [0.2cm]
 $r_{\rm s,min}$ [kpc] & $0.11^{+0.02}_{-0.04}$ & $0.11^{+0.02}_{-0.04}$ & $0.11^{+0.02}_{-0.04}$ \\ [0.2cm]
\hline
\end{tabular}\par
Table A2.2
\end{minipage}

\subsubsection{z=1}

\begin{minipage}{\linewidth}
  \footnotesize
    \bigskip
 \centering
 \begin{tabular}{ | c | c | c | c |}
\hline
 & $m_{\rm high}=10^8$ M$_{\odot}$    & $m_{\rm high} = 10^9$ M$_{\odot}$  & All subhalos  \\ \hline
 $N_{\rm sub}$($L_{\rm box}=300$ kpc)  & 462  & 511 & 521  \\ [0.2cm]
 $N_{\rm sub}$($L_{\rm box} = 100$ kpc) & $56^{+14}_{-14}$ & $62^{+12}_{-14}$ & $65^{+11}_{-20}$ \\ [0.2cm]
 $N_{\rm sub}(R_{\rm Ein}$) & $1^{+3}_{-1}$ & $1^{+3}_{-1}$ & $1^{+3}_{-1}$ \\ [0.2cm]
 $\bar{\kappa}_{\rm sub}$ & $\left(1.17^{+0.55}_{-0.44}\right) \times 10^{-5}$ & $\left(4.68^{+2.68}_{-2.74}\right) \times 10^{-5}$ & $\left(9.58^{+11.30}_{-6.86}\right)\times 10^{-5}$ \\ [0.2cm]
 $\langle m \rangle$ [M$_{\odot}$] & $\left(9.33^{+2.87}_{-2.75}\right) \times 10^6$ & $\left(3.33^{+2.28}_{-1.59}\right) \times 10^7$ & $\left(6.26^{+10.10}_{-3.81}\right) \times 10^7$ \\ [0.2cm]
 $m_{\rm eff} \equiv \langle m^2 \rangle / \langle m \rangle$ [M$_{\odot}]$ & $\left(3.32^{+1.11}_{-1.75}\right) \times 10^7$ & $\left(2.71^{+2.41}_{-1.70}\right) \times 10^8$ & $\left(7.84^{+49.50}_{-5.85}\right) \times 10^8$ \\ [0.2cm]
 $r_{\rm t,max}$ [kpc]& $14.17^{+21.39}_{-5.24}$ & $22.73^{+22.16}_{-11.53}$  & $23.11^{+21.78}_{-9.12}$ \\ [0.2cm]
 $r_{\rm s,min}$ [kpc] & $0.12^{+0.04}_{-0.04}$  & $0.12^{+0.04}_{-0.04}$ & $0.12^{+0.04}_{-0.02}$  \\ [0.2cm]
\hline
\end{tabular}\par
Table A2.3
\end{minipage}
\end{widetext}

\bibliography{sample}
\bibliographystyle{apsrev4-1}

\end{document}